\documentclass[fleqn]{cas-dc}

\usepackage{amsmath}
\usepackage{mathtools}
\usepackage{url}
\usepackage{gensymb}
\usepackage[authoryear]{natbib}
\usepackage{lineno}

\let\oldequation\equation
\let\oldendequation\endequation

\renewenvironment{equation}
  {\linenomathNonumbers\oldequation}
  {\oldendequation\endlinenomath}

\usepackage{threeparttable}

\def\tsc#1{\csdef{#1}{\textsc{\lowercase{#1}}\xspace}}
\tsc{WGM}
\tsc{QE}
\tsc{EP}
\tsc{PMS}
\tsc{BEC}
\tsc{DE}
\newcommand{\de}{{\rm d}}

\begin{document}
\let\WriteBookmarks\relax
\def\floatpagepagefraction{1}
\def\textpagefraction{.001}

\shorttitle{Dark Comet Origins}

\shortauthors{Taylor et al.}

\title [mode = title]{The Dynamical Origins of the Dark Comets and a Proposed Evolutionary Track}                      

\author[1]{Aster G. Taylor}[orcid=0000-0002-0140-4475]

\cormark[1]

\fnmark[1]

\ead{agtaylor@umich.edu}

\affiliation[1]{organization={Dept. of Astronomy, University of Michigan},
    city={Ann Arbor},
    postcode={48109}, 
    state={MI},
    country={USA}}

\author[2,3]{Jordan K. Steckloff}[orcid=0000-0002-1717-2226]
\affiliation[2]{organization={The Planetary Science Institute},
    city={Tucson},
    state={AZ},
    country={USA}}

\affiliation[3]{organization={Dept. of Aerospace Engineering and Engineering Mechanics, The University of Texas at Austin},
    city={Austin},
    state={TX},
    country={USA}}

\author[4]{Darryl Z. Seligman}[orcid=0000-0002-0726-6480]
\fnmark[2]
\affiliation[4]{organization={Dept. of Astronomy and Carl Sagan Institute, Cornell University},
    addressline={122 Sciences Drive}, 
    city={Ithaca},
    postcode={14853}, 
    state={NY},
    country={USA}}

\author[5]{Davide Farnocchia}[orcid=0000-0003-0774-884X]
\affiliation[5]{organization={Jet Propulsion Laboratory, California Institute of Technology},
    addressline={4800 Oak Grove Dr.}, 
    city={Pasadena},
    postcode={91109}, 
    state={CA},
    country={USA}}

\author[6]{Luke Dones}[orcid=0000-0001-8573-7412]
\affiliation[6]{organization={Southwest Research Institute},
    addressline={1050 Walnut Street, Suite 300}, 
    city={Boulder},
    postcode={80302}, 
    state={CO},
    country={USA}}

\author[7]{David Vokrouhlick\'y}[orcid=0000-0002-6034-5452]
\affiliation[7]{organization={Institute of Astronomy, Charles University},
    addressline={V Hole\v{s}ovi\v{c}k\'ach 2, CZ-18000 }, 
    city={Prague},
    citysep={}, 
    postcode={8}, 
    country={Czech Republic}}

\author[6]{David Nesvorn\'y}[orcid=0000-0002-4547-4301]

\author[8]{Marco Micheli}[orcid=0000-0001-7895-8209]
\affiliation[8]{organization={ESA NEO Coordination Centre},
    addressline={Largo Galileo Galilei 1, I-00044 Frascati (RM)}, 
    country={Italy}}

\cortext[cor1]{Corresponding author}

\fntext[fn1]{Fannie and John Hertz Foundation Fellow}
\fntext[fn2]{NSF Astronomy and Astrophysics Postdoctoral Fellow}

\begin{abstract}
So-called `dark comets' are small, morphologically inactive near-Earth objects (NEOs) that exhibit nongravitational accelerations inconsistent with radiative effects. These objects exhibit short rotational periods (minutes to  hours), where measured. We find that the strengths required to prevent catastrophic disintegration are consistent with those measured in cometary nuclei and expected in rubble pile objects. We hypothesize that these dark comets are the end result of a rotational fragmentation cascade, which is consistent with their measured physical properties. We calculate the predicted size-frequency distribution for objects evolving under this model. Using dynamical simulations, we further demonstrate that the majority of these bodies originated from the $\nu_6$ resonance, implying the existence of volatiles in the current inner main belt. Moreover, one of the dark comets, (523599) 2003 RM, likely originated from the outer main belt, although a JFC origin is also plausible. These results provide strong evidence that volatiles from a reservoir in the inner main belt are present in the near-Earth environment. 
\end{abstract}



\maketitle

\section{Introduction} \label{sec:intro}

Classically, small solar-system bodies are divided into two broad populations --- asteroids and comets. In general, asteroids are on short-period orbits and lack visibly sublimating volatiles. The previously-measured nongravitational accelerations on asteroids are generally attributed to radiation pressure \citep{Vokrouhlicky2000} or, in most cases, the Yarkovsky effect \citep{Farnocchia2013, Vokrouhlicky2015, Greenberg2020}. Most asteroids have typical densities of $\rho\sim1000-3000$~kg/m$^{3}$ \citep{Carry2012} and reside in the main belt between Mars and Jupiter or in the Trojan populations co-orbiting with Jupiter. Due to the Yarkovsky effect, some asteroids drift into resonances such as the $\nu_6$ secular resonance and the 3:1 mean-motion resonance with Jupiter \citep{Gladman1997, Migliorini1998, Granvik2017}. These resonances excite the asteroids' eccentricities, causing many of them to dynamically evolve into near-Earth objects (NEOs) with perihelia of $q\leq 1.3$ au \citep{Nesvorny2017, Seligman2021, Nesvorny2023}. However, the near-Earth environment is dynamically unstable and NEOs are continuously removed via scattering into the Sun, scattering onto Jupiter-crossing orbits, and planetary collisions \citep{Farinella1994, Nesvorny2023}.

On the other hand, most comets are thought to originate from either the Oort Cloud or the Trans-Neptunian Object (TNO) population, where low ambient temperatures allow volatiles to exist for extended periods \citep{Volk2008,Nesvorny2018,Lisse2022b}. The presence of relatively light volatiles and high nuclear porosity in comets results in consistently lower bulk densities than asteroids, typically on the order of $\rho\sim500$~kg/m$^{3}$ \citep{Knight2023}. Comets typically exhibit their eponymous comae and nongravitational accelerations, which arise from the outgassing of sublimating volatiles upon entering the inner solar system \citep{Whipple1950,Whipple1951,Yeomans2004Comets2}. Recoil forces from outgassing tend to spin up these nuclei, resulting in rotational fragmentation as centrifugal forces overcome the binding forces \citep{Steckloff2016a, Jewitt2021, Jewitt2022dest}. This effect, in combination with mass loss due to outgassing, generally leads to ``cometary fading'', the destruction of comets on relatively short timescales \citep{Wiegert1999, Brasser2015}.  

It is expected that most short-period comets, particularly Jupiter Family and Encke-type comets, originate from the TNO reservoir \citep{Levison1997, Nesvorny2018}. Analogously to the main belt, interactions with the giant planets (particularly Neptune) can destabilize and scatter the orbits of Kuiper Belt objects (KBOs) inwards \citep{Nesvorny2017}. These KBOs then transition to the giant planet region between Jupiter and Neptune. At this point, these objects are classified as Centaurs, named after the first object {recognized to be} in this class, (2060) Chiron \citep{Kowal1979}. While definitions vary, Centaurs are generally understood to be objects with semimajor axes of $a\geq5$ au and perihelia of $q\leq30$ au \citep{Jewitt2009}. 

Centaur orbits are unstable and chaotically evolve due to their proximity to the giant planets. About one-third of Centaurs are scattered inwards via interactions with Jupiter and become JFCs after $\sim1-10$ Myr as a Centaur \citep{Tiscareno2003, DiSisto2007, Seligman2021}. JFCs are then destroyed by repeated perihelion passages \citep{Nesvorny2018}, ejected from the solar system, or dynamically detached from Jupiter via migration into the near-Earth environment \citep{Grav2023,Nesvorny2023}.

Recent developments have demonstrated that a continuum of objects exists between the two extremes of asteroids and comets, such as the active asteroids and inactive comets. Active asteroids are objects on asteroid-like orbits with observable cometary activity \citep{Jewitt2012, Hsieh2017}. A particular example is the Main Belt Comets (MBCs), which reside within the main belt but exhibit comae \citep{Hsieh2006}. {It is worth mentioning that the MBCs are a subset of active asteroids. MBC activity is due to sublimation of ice, while active asteroids may be active due to rotational instability or impacts \citep{Jewitt2012}.} Only a handful of objects have been identified as MBCs, and targeted searches suggest occurrence rates of $<$1/500 and $\sim$1/300 in the outer main belt \citep{Sonnett2011, Bertini2011, Ferellec2023}. See \citet{Jewitt2022} for a recent review. 

Inactive comets are on cometary orbits but display no detectable activity. While there are several categories of inactive comets, classified based on their orbits, nearly all are likely extinct or dormant comet nuclei that were not destroyed by outgassing or outgassing-driven processes \citep{Asher1994, Jewitt2005, Licandro2018}. There are also a few known weakly active, tailless ``Manx'' comets \citep{Meech2016, Piro2021, Kwon2022}. It has been suggested that the Manx comets are asteroids that were scattered into the Oort cloud and developed a volatile-rich frost before traveling back toward the Sun \citep{Weissman1997, Shannon2015, Meech2016}. 

Recently, \citet{Farnocchia2023} and \citet{Seligman2023} \citep[building on the work of][]{Chesley2016} defined ``dark comets'' as a population of active asteroids that are distinct from these previously identified categories. The dark comets are on near-Earth orbits and exhibit no detectable comae (hence the ``dark'' {moniker}), yet experience nongravitational accelerations that are inconsistent with radiative effects. \citet{Farnocchia2023} and \citet{Seligman2023} demonstrated that outgassing is consistent with both the reported accelerations and the nondetections of comae. In addition to their accelerations and absence of comae, dark comets exhibit several other notable properties --- they are generally (i)  small (on the order of $10-100$ m), (ii) rapidly rotating \citep[{$0.046-1.99$ h,}][]{Seligman2023}, and (iii) accelerating in nonradial directions. 

\citet{Taylor2024} demonstrated that if rapid rotation causes diurnal outgassing to be negligible, differential heating across the hemispheres of these objects due to seasonal effects could generate the observed out-of-plane acceleration. This mechanism is consistent with the accelerations measured on all the currently known dark comets except for 2003 RM, which also differs from the other objects in size and orbital parameters. {However, this mechanism relies on the rotation rate, which is unmeasured on a majority of the dark comets.}

Here, we further investigate the properties and origins of the dark comets. This paper is organized as follows: in Sec.~\ref{sec:popest}, we {calculate a preliminary estimate of the} population fraction in the current near-Earth environment. In Sec.~\ref{sec:matstr}, we calculate the minimum material strength required to hold the dark comets stable against rotational disruption, and compare with the estimated strengths of known comet nuclei. In Sec.~\ref{sec:evopath}, we introduce a proposed evolutionary track for these objects, in which rotational splitting and fragmentation results in nuclei consistent with dark comets. In Sec.~\ref{sec:sfd}, we calculate the size-frequency distribution that our model predicts for dark comets. In Sec.~\ref{sec:dynorig}, we discuss the dynamical transfer of objects to the dark comet orbits. Finally, in Sec.~\ref{sec:discussion}, we discuss the implications of our results. 

\section{Approximate Occurrence Rate}\label{sec:popest}

Given the long temporal baselines of astrometric data required to measure nongravitational accelerations and the small sizes of the dark comets, it is possible that more as-yet undetected dark comets exist in the inner solar system. In this section, we provide an order-of-magnitude estimate of the population of dark comets, {albeit one subject to large uncertainties due to small-number statistics, unknown detection biases, and poorly constrained evolutionary histories}. 

To {estimate} the fraction of NEOs that are dark comets, we first estimate the fraction of NEOs that can be ruled out as dark comets. Specifically, we calculate a best-fit nongravitational acceleration for each NEO, including uncertainty. We then find the number of objects with sufficiently small uncertainties in hypothetical nongravitational accelerations to rule out dark comet--like values. The population-level occurrence rate is then approximately the number of {known} dark comets divided by the total number of objects within this subset. As only 7 dark comets are currently known, this is a first-order estimate. 

We estimated nongravitational accelerations for 10\,553 NEOs using the least-squares fitting method of \citet{Farnocchia2015}. {Notably, the same methodology was used to identify the dark comets, thereby cancelling any biases in the estimation of the nongravitational accelerations}. The optical astrometric data are obtained from the Minor Planet Center\footnote{https://www.minorplanetcenter.net/} and radar measurements from the Solar System Dynamics Group\footnote{https://ssd.jpl.nasa.gov/sb/radar.html} at JPL. The nongravitational accelerations are modeled as $A_i\,(\text{1 au}/r)^2$ in each of the {orbital} coordinate directions (radial, transverse, out-of-plane), following the model of \citet{MarsdenII}. 

Many of these objects will have nongravitational accelerations from radiation pressure or the Yarkovsky effect, independent of any dark comet--like behavior. 
{To account for this, we} obtain best-fit lines for the acceleration magnitude of radiation pressure and the Yarkovsky effect, with object data given by the JPL Small-Body Database\footnote{https://ssd.jpl.nasa.gov/tools/sbdb\_query.html}. 
We assume that radiation pressure and the Yarkovsky effect both exhibit a $1/R_{\rm N}$ dependence, where $R_{\rm N}$ is the radius of the nucleus and is computed from the absolute magnitude $H$ by 
\begin{equation}\label{eq:RofH}
    R_{\rm N} = \bigg(\frac{1329\text{ km}}{2\sqrt{p}}\bigg)10^{-0.2H}\,.
\end{equation}
Eq.~\eqref{eq:RofH} is given by \citet{Pravec2007}. The expected radiative nongravitational accelerations are given by 
\begin{equation}\label{eq:Aest}
    A_{i, {\rm est}}=10^{0.2H+b_i}\text{ au/yr$^2$}\,.
\end{equation}
In Eq.~\eqref{eq:Aest}, $b_i$ is a constant scaling factor, which is {$-10.734\pm0.075$} for radiation pressure ($i=1$) and {$-12.315\pm0.014$} for the Yarkovsky effect \citep[$i=2$; see][]{Seligman2023}.\footnote{{These uncertainties are small compared to the parameter values and are a minimal source of error.}} {Note that this equation depends only on the absolute magnitude $H$ and the empirically-determined $b_i$ and has no dependence on the albedo. Therefore, this estimate of the occurrence rate holds for any subpopulation of asteroids with a representative albedo.} These values are empirically obtained as the best-fit line to object data. Although there is no known out-of-plane radiative effect, we assume that the expected out-of-plane nongravitational acceleration ($i=3$) is equivalent to the Yarkovsky effect {as a proxy}. 

\begin{table*}[t]
\begin{center}   
\caption{\textbf{Dark Comet Nongravitational Accelerations.} {The nongravitational accelerations' magnitudes, significance, and the maximum ratio of measured nongravitational acceleration to expected nongravitational acceleration for all of the dark comets. Nongravitational accelerations and uncertainties are given by \citet{Seligman2023} and the references therein.} } 
\begin{threeparttable}
\begin{tabular}{l r@{$\pm$}l r r@{$\pm$}l r r@{$\pm$}l r r } \label{table:objaccs}
 Object & \multicolumn{2}{c}{$A_1$\tnote{a}} & Signif.\tnote{b} & \multicolumn{2}{c}{$A_2$\tnote{c}} & Signif.\tnote{d} & \multicolumn{2}{c}{$A_3$\tnote{e}} & Signif.\tnote{f} & max($C_i$)\tnote{g} \\
 & \multicolumn{2}{c}{[$10^{-5}$ au yr$^{-2}$]} & [$\sigma$] & \multicolumn{2}{c}{[$10^{-5}$ au yr$^{-2}$]} & [$\sigma$] & \multicolumn{2}{c}{[$10^{-5}$ au yr$^{-2}$]} & [$\sigma$] & \\
 \hline
1998 KY$_{26}$ & 2.31&1.21 & 2 & -0.00168&0.00081 & 2 & 0.426&0.153 & 3 & 66.86 \\
2005 VL$_1$ & -8.87&10.7 & $<$1 & -0.00947&0.00789 & 1 & -0.320&0.0546 & 6 & 33.90 \\
2016 NJ$_{33}$ & 12.4&3.94 & 3 & -0.00754&0.00257 & 3 & 1.13&0.217 & 5 & 186.38 \\
2010 VL$_{65}$ & 8.75&17.3 & $<$1 & -0.00195&0.00711 & $<$1 & -1.22&0.173 & 7 & 36.01 \\ 
2010 RF$_{12}$ & 0.650&0.795 & $<$1 & -0.00181&0.00381 & $<$1 & -0.224&0.028 & 8 & 9.58 \\ 
2006 RH$_{120}$ & 1.84&0.107 & 18 & -0.676&0.0849 & 8 & -0.173&0.0426 & 4 & 17.58 \\
2003 RM & -1.39&1.62 & $<$1 & 0.0286&0.0005 & 56 & 0.0200&0.0723 & $<$1 & 679.90 \\\hline
\end{tabular}
\begin{tablenotes}[para]
    \item[a]{{Radial nongravitational acceleration.}}
    \item[b]{{Significance of $A_1$.}}
    \item[c]{{Transverse nongravitational acceleration.}}
    \item[d]{{Significance of $A_2$.}}
    \item[e]{{Out-of-plane nongravitational acceleration.}}
    \item[f]{{Significance of $A_3$.}}
    \item[g]{{Maximum value of the ratio of the measured to expected nongravitational acceleration across all three components.}}
\end{tablenotes}
\end{threeparttable}
\end{center}
\end{table*}

\begin{figure}
    \centering
    \includegraphics{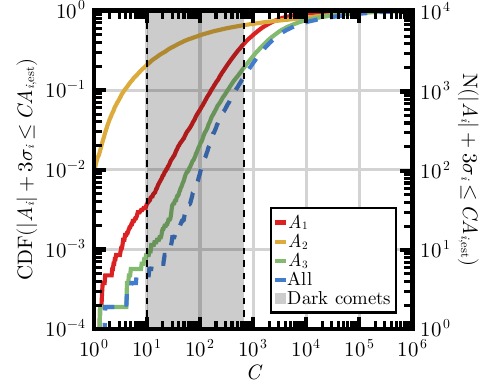}
    \caption{Cumulative fraction/number of objects with nongravitational accelerations smaller than a given cutoff, which is a factor of $C$ larger than the estimated nongravitational acceleration from radiation pressure ($i=1$) or the Yarkovsky effect ($i=2$, and $i=3$ by proxy). All three components are shown individually. The number of objects that satisfy Eq.~\eqref{eq:Aicond} for all 3 components are also plotted. The known dark comets have accelerations $\sim${10 -- 680}$\times$ the expected radiative values, which are shown as a {shaded region bounded by} vertical dashed lines. }
    \label{fig:CDFest}
\end{figure}

We define a cutoff value $C A_{i, {\rm est}}$, signifying the value where nongravitational accelerations are considered to be significant in comparison to typical values {from radiation-driven effects. For example, an object with a radius of $R\simeq 100$ m {and an albedo of $p=0.1$} is expected to have a radiation-pressure acceleration of $A_{1,{\rm est}}\sim4\times10^{-7}$ au yr$^{-2}$}. For a given $C$, we count the number of objects which have nongravitational acceleration magnitudes smaller than this cutoff by $>3\sigma$. Therefore, the estimated fraction of objects with nongravitational accelerations definitively lower than $C A_{i, {\rm est}}$ is given by the fraction of objects that satisfy
\begin{equation}\label{eq:Aicond}
    |A_i|+3\sigma_i\leq C A_{i, {\rm est}}\,.
\end{equation}
In Eq.~\eqref{eq:Aicond}, $\sigma_i$ is the uncertainty in the $i$th nongravitational acceleration component. In Fig.~\ref{fig:CDFest}, we plot the number/fraction\footnote{The total number is the same for all three components of the nongravitational acceleration by construction of our sample.} of objects that satisfy this condition for a range of values of $C$. The dark comet accelerations are {$\sim$10 -- 680$\times$} the expected radiative values {(Table~\ref{table:objaccs})}, {providing an upper and lower bound of the estimated fraction of objects that can be ruled out as dark comets. These boundaries depend on the typical magnitude of a dark comet nongravitational acceleration, and will be further constrained with future discoveries.} We find that for $C=10$ -- $680$, {a range of} 5 -- 1561 objects satisfy Eq.~\eqref{eq:Aicond} for all 3 acceleration components simultaneously.\footnote{{Note that since dark comets are identified by a detection in at least one of the acceleration components, ruling out an object requires a nondetection in all three acceleration components.}} Therefore, this population represents the subset that can be ruled out as dark comets {via this method}. Since 7 dark comets are currently known, we estimate that 0.5\% -- 60\% of all sufficiently-constrained NEOs are dark comets. It is important to note that this estimate is subject to small-number statistics and heretofore uncharacterized observational biases. {As a result, it is not currently possible to accurately estimate the occurrence rate of dark comets from given data. Future data are necessary to refine this estimate. }

If this occurrence rate is representative of the entire NEO population, it is possible that a significant fraction of all NEOs, or tens to thousands of objects, could be dark or dead comets. However, our current lack of understanding of the history, origins, and observational constraints on the dark comets makes an exact calculation {impossible}. The small sizes of the dark comets also restrict observations to objects with close approaches to Earth and long data arcs, which may not reflect the overall NEO population. As data arcs extend and new dark comets are discovered, the known population will increase and population estimates will become more robust.

\section{Material Strengths}\label{sec:matstr}

\begin{figure}
    \centering
    \includegraphics[width=\linewidth]{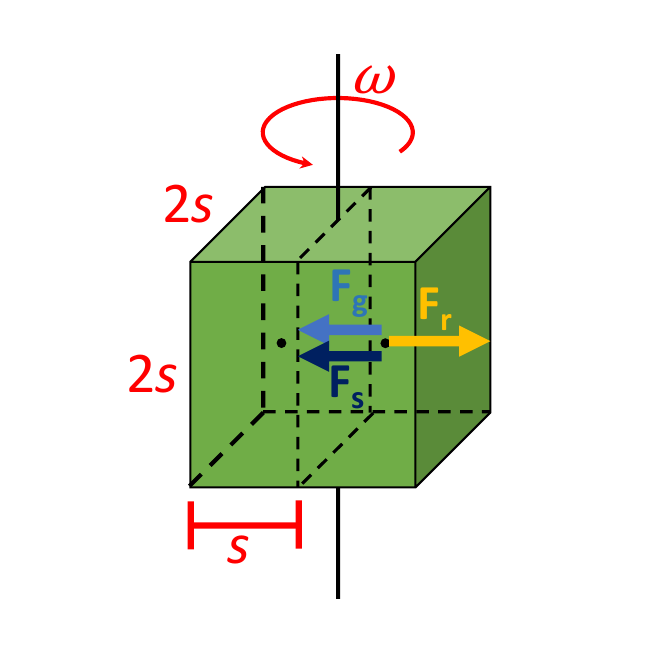}
    \caption{A diagram of our model of a cometary nucleus. We represent a simplified nucleus as a cube with a side length of $2s$. The cube is composed of two identical rectangular prisms with sides of length $s$, $2s$, $2s$. The rotation axis passes through the center of this structure. The relevant destabilizing and binding forces are shown in orange and blue, respectively.}
    \label{fig:cartoon}
\end{figure}

In this section, we calculate the minimum tensile strength required to stabilize dark comets against rotational disruption (a basic property of geologic materials). We construct a simple three-force model comparing the centripetal force with two binding forces (material strength and self-gravity) required to hold the rotating object together. We approximate the nucleus as a cube of side length $2s$, with the rotation axis passing through the center of two opposing faces. With this approximation, the cube can be divided into two rectangular prisms joined along a plane passing through the rotation axis, each with dimensions of $s$:$2s$:$2s$ (Fig.~\ref{fig:cartoon}). 

While previous work has calculated the internal stress tensors and critical rotation rates for triaxial ellipsoids \citep{Davidsson1999, Davidsson2001, Breiter2012}, this model allows for an intuitive understanding of the relevant forces, while being sufficient to obtain the tensile strength with errors of order unity. In addition, observational uncertainties dominate the error that arises from our model choice, so improving this model would not necessarily produce a more reliable strength measurement. Our cubic nucleus approximation also has a smaller aspect ratio than many comet nuclei \citep{Knight2023}. The resulting tensile strengths that we calculate are therefore biased downwards, {so our calculations} provide a minimum strength value. However, we can find a similar constraint by modeling the nucleus as a {2:1} rectangular prism comprised of two cubes, which produces nearly-identical estimates of the material strength. {The aspect ratio is therefore not a significant factor in the material strength of these objects. Note that only 1998 KY$_{26}$ has a measured shape model, which is roughly spherical \citep{Ostro1999}. The rapid rotation rates of 2016 NJ$_{33}$ and 2006 RH$_{120}$ imply that these objects likely do not have extreme aspect ratios that make them more susceptible to rotational disruption.}

Tensile forces on a homogeneous rotating object of uniform cross-section are maximized along a plane passing through the rotation axis. In the case of fragmentation, we assume the object materially fails and produces the two rectangular prisms as child nuclei.\footnote{Although this binary division is simplistic, it simplifies the force calculations while introducing only minor errors.} The strength of the gravitational binding force is  the force resulting from two identical rectangular prisms of side length $s$, $2s$, and $2s$ and density $\rho$ in contact with one another such that their centers are separated by a distance $s$. The physical binding force $F_{\rm s}$ of the nucleus is the product of the contact area of the two cubes and their tensile strength $\sigma_t$. The gravitational force is $Gm^2/s^2$ ($G$ is the gravitational constant), the tensile strength force is $\sigma_t(2s)^2$, and the mass of each prism is $m=4\rho s^3$. The binding force is therefore given by
\begin{equation}\label{eq:Fbinding}
    F_{\rm bind} = 16G\rho^2 s^4 +4 \sigma_{t}s^2\,.
\end{equation}

For a given rotation rate, the centripetal force needed to keep a mass of $m$ in a circle at a distance of $s/2$ from the axis (see Fig.~\ref{fig:cartoon}) is given by
\begin{equation}\label{eq:Fcentripetal}
    F_{\rm r} = 2\rho s^4 \omega^2.
\end{equation}
In Eq.~\eqref{eq:Fcentripetal}, $\omega$ is the angular velocity, which is related to the rotation period $P$ by $P=2\pi/\omega$. These objects are minimally stable against rotational disruption when this centripetal force balances the binding forces. The force required at this minimum strength is therefore a lower limit on the bulk tensile strength of these dark comets. We set Eqs.~\eqref{eq:Fbinding} and \eqref{eq:Fcentripetal} equal and rearrange to find that the minimum strength is given by
\begin{equation}\label{eq:sigmatdef}
\sigma_{t} = \rho s^2 \left(\frac{1}{2}\omega^2 - 4G\rho\right)\,.
\end{equation}
As most of the dark comets have been observed for a decade or more, we assume they are currently stable against rotational disruption.

\begin{table*}[t]
\caption{\textbf{Dark Comets.} Orbital parameters, absolute magnitudes, sizes, rotation periods, and estimated limits on material strengths of the dark comets. {Parameters are given by the JPL Horizons database (\url{https://ssd.jpl.nasa.gov/horizons.cgi})}. Sizes are given by \citet{Seligman2023} and the references therein. The rotational period of 1998 KY$_{26}$ was reported by \citet{Ostro1999}, the period of 2016 NJ$_{33}$ was reported by \citet{Seligman2023}, and the period of 2006 RH$_{120}$ was reported by \citet{Kwiatkowski2009}. The remaining objects do not have reported measurements of rotational period. } 
\label{table:objects}
\begin{threeparttable}
\begin{tabular}{lcccccccc }
 Object & $a$\tnote{a} & $e$\tnote{b} & $i$\tnote{c} & H\tnote{d} & $R$\tnote{e} & $P$\tnote{f} & $\sigma_{t,c}>$\tnote{g} & $\sigma_{t,a}>$\tnote{h} \\
 & [au] & & [$^\circ$] & [mag] & [m] & [h] & [Pa] & [Pa] \\
 \hline
1998 KY$_{26}$ & 1.23 & 0.20 & 1.48 &  25.60 & 15& 0.178 & 5.3 & 26 \\
2005 VL$_1$ & 0.89 & 0.23 & 0.25 & 26.45  & 11 & &  & \\
2016 NJ$_{33}$ & 1.31 & 0.21 & 6.64 & 25.49 & 16 & $0.41$-$1.99$ & $0.0$-$1.1$ & $0.0$-$5.6$ \\
2010 VL$_{65}$ & 1.07 & 0.14 & 4.41 & 29.22 & 3& & & \\ 
2010 RF$_{12}$ & 1.06 & 0.19 & 0.88 & 28.42 &4 & &  & \\ 
2006 RH$_{120}$& 1.00 & 0.04 & 0.31 & 29.50 & 2-7 & 0.046 & 1.4-18 & 7.5-91 \\
2003 RM & 2.92 & 0.60 & 10.86 & 19.70 & 230 &  &  & \\\hline
\end{tabular}
\begin{tablenotes}[para]
    \item[a]{Semimajor axis.}
    \item[b]{Eccentricity.}
    \item[c]{Inclination.}
    \item[d]{Absolute magnitude.}
    \item[e]{Nuclear radius.}
    \item[f]{Rotation period.}
    \item[g]{Minimum material strength for a cometary density of $\rho\simeq 500$ kg/m$^{3}$.}
    \item[h]{Minimum material strength for an asteroidal density of $\rho\simeq 2600$ kg/m$^{3}$.}
\end{tablenotes}
\end{threeparttable}
\end{table*}

Using Eq.~\eqref{eq:sigmatdef}, we compute this lower limit for the dark comet tensile strengths from their measured rotational periods (currently measured for three dark comets). We assume that the length scale $s$ is equal to the estimated radius for each. For each dark comet with a known rotational period (1998 KY$_{26}$, 2016 NJ$_{33}$, and 2006 RH$_{120}$), we provide a range of minimum strength values, based on uncertainty in the periods and sizes. We calculate these tensile strengths assuming both (i) a typical comet nucleus density \citep[500 kg/m$^{3}$,][]{Richardson2007,Patzold2016} and (ii) a typical chondrite meteorite density \citep[2600 kg/m$^{3}$,][]{Yeomans2000}. These results are presented in Table~\ref{table:objects} as $\sigma_{t,c}>$ and $\sigma_{t,a}>$ respectively.

These computed minimum strengths are on the order of $\sigma_{t,c}\gtrsim 0.01$ -- $18$ Pa or $\sigma_{t,a}\gtrsim 0.0$ -- $91$ Pa, assuming the objects have a typical comet density or typical asteroid density, respectively. These strengths are consistent with those of comet nuclei, which are on the order of $\sigma_t\sim1$--$10$ Pa \citep{Sekanina1985,Asphaug1996,Steckloff2015,Attree2018}, with other estimates reaching as high as $\sigma_t\sim10-200$ Pa for 67P \citep{Hirabayashi2016}. Even for rubble pile bodies held together purely by van der Waals interactions between grains, tensile strengths are on the order of a few tens of Pa \citep{Sanchez2014,Rozitis2014}, such as for the \textit{DART} target's primary Didymos \citep{Zhang2017, Zhang2021}. Therefore, these strengths are entirely consistent with rubble-pile bodies, {and no additional cohesion is required}.

However, dark comets are much smaller than typical {cometary} nuclei. Because the binding force terms have different dependencies on the size $s$ of the object, dark comets may exhibit fundamentally different evolutionary behavior if they are predominantly bound by a different force than the larger {cometary} nuclei. This is evidenced in the ``spin barrier'' for larger comets and asteroids, in which rubble piles (gravity-dominated objects) break apart at rotation periods shorter than $\sim2.2$ hours for asteroids \citep{Pravec2002, Warner2009,  Sanchez2014} or $\sim6$ hours for comets \citep{Kokotanekova2017,Safrit2021}. However, smaller, strength-dominated objects can rotate more rapidly if they have significant material strength.

We estimate the approximate size at which objects transition between strength- and gravity-dominated regimes. To do this we calculate the angular velocity at which the rotational forces overcome the tensile strength and gravitational forces and set these velocities equal to each other. 

\begin{figure}[t]
    \centering
    \includegraphics{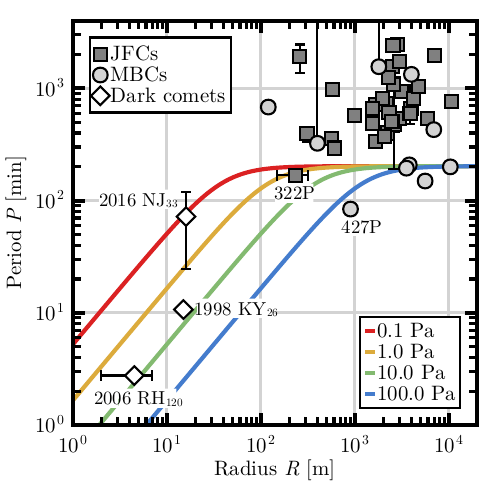}
    \caption{Minimum stable rotation period (fastest stable rotation rate) as a function of size and tensile strength. Here we use our simple cubic model to compute the binding forces (self-gravity and strength) to determine the size-dependence of the minimum stable rotation period for various tensile strengths. {Note that a larger aspect ratio would increase the tensile strength required to maintain the cohesion of the body.} We find that the strength of a typical comet nucleus \citep[$\sim$1--100 Pa; ][]{Sekanina1985,Asphaug1996,Bowling2014,Steckloff2015,Hirabayashi2016,Attree2018} can hold small comet nuclei together at rotation rates faster than those of the dark comets. Therefore, dark comets do not require atypical strengths, in spite of their short rotation periods/fast rotation rates. The three dark comets with measured rotation periods are shown as white diamonds. JFCs from \citet{Kokotanekova2017} are shown as {dark gray} squares, {and MBCs from \citet{Jewitt2022} are shown as light gray circles}. As the radius gets larger, gravity begins to dominate over strength, so the minimum stable period converges.}
    \label{fig:size-period}
\end{figure}

The rotational failure criterion for a strength-dominated object is given by
\begin{equation}\label{eq:wcrit}
\omega_{\rm crit}^{\rm str} = \sqrt{\frac{2 \sigma_t}{\rho s^2}}\,.
\end{equation}
For a derivation of this relationship, see \citet{Steckloff2016a}. Similarly, the failure criterion for a gravity-dominated object is 
\begin{equation}
    \omega_{\rm crit}^{\rm grav} = \sqrt{\frac{4\pi\rho G}{3}}\,.
\end{equation}
Notably, this is independent of object radius \citep{Pravec2000,Safrit2021}{, although it will depend on the object's aspect ratio}. Setting these two equations equal to one another provides an expression for the strength-to-gravity transition size --- $R\sim200$ m for asteroids and $R\sim1$ km for comets (assuming tensile strength of 10 Pa and densities of 2600 kg/m$^{3}$ and 500 kg/m$^{3}$ respectively). This transition size can be scaled to different densities and tensile strengths, such that
\begin{equation}
    s_{\rm trans}^{\rm ast} = 206\text{ m } \left(\frac{\sigma_t}{10 \text{ Pa}}\right)^{1/2}\left(\frac{\rho}{2600 \text{ kg m}^{-3}}\right)^{-1}\,.
\end{equation}
Similarly, the transition size for comets is 
\begin{equation}
    s_{\rm trans}^{\rm comet} = 1070\text{ m } \left(\frac{\sigma_t}{10 \text{ Pa}}\right)^{1/2}\left(\frac{\rho}{500 \text{ kg m}^{-3}}\right)^{-1}\,.
\end{equation}
This asteroid transition radius is consistent with that calculated by \citet{Pravec2002}. This suggests that all the dark comets are small enough to lie in the strength-dominated regime {(Fig.~\ref{fig:size-period})}, in contrast to the comet nuclei visited by spacecraft, all of which are large enough to be gravity-dominated: 9P/Tempel 1 \citep{A'Hearn2005}, 19P/Borrelly \citep{Soderblom2002}, 67P/Churyumov-Gerasimenko \citep{Jorda2016}, and 81P/Wild 2 \citep{Brownlee2006}, with 103P/Hartley 2 \citep[$s \sim 600$~m;][]{A'Hearn2011} possibly representing a transition object.

\begin{figure*}[t]
    \centering
    \includegraphics[width=\linewidth]{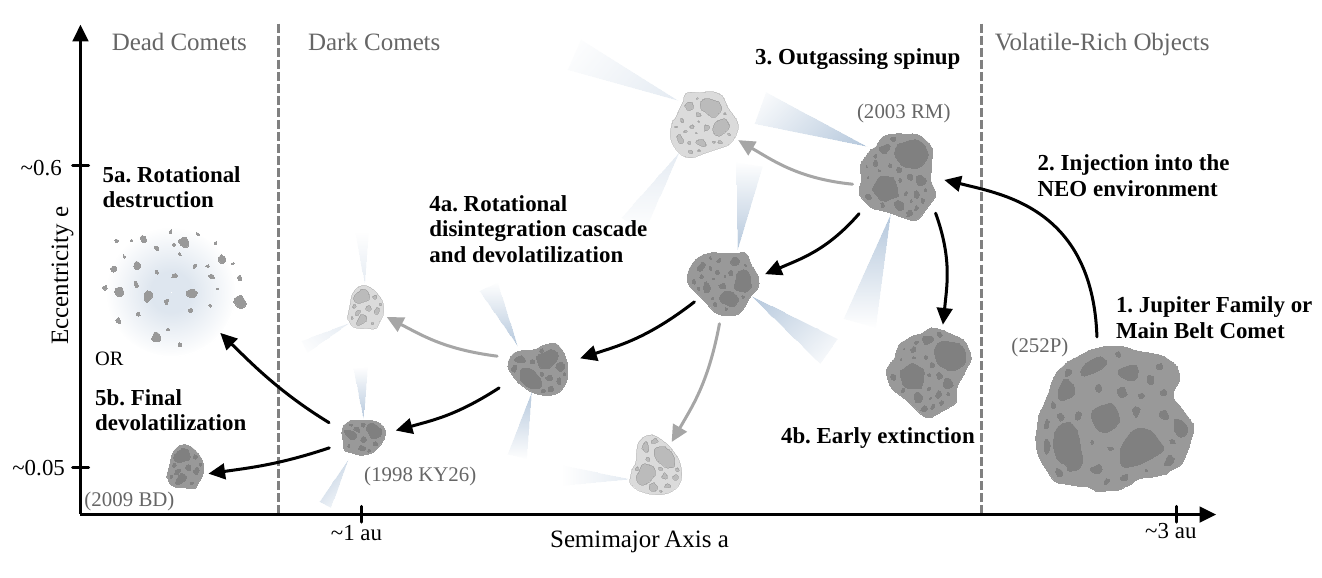}
    \caption{The proposed evolutionary track for the dark comets --- a {comet} is injected into the near-Earth environment, where outgassing causes spinup. The spinup causes a rotational disintegration cascade and devolatilization, leading to smaller, less active objects (i.e., dark comets). {Extinction is also possible at this point in the evolutionary track, well before becoming an NEO (4b).} If the disintegration cascade operates more rapidly than devolatilization (Sec.~\ref{subsec:devoltime}), then the object will eventually be destroyed completely. However, if volatiles are removed before this can take place, then the end state will be a small, rapidly-rotating, inactive asteroid in the near-Earth environment. {Example objects are also given, which represent analogous physical (not dynamical) stages of this model.}  }
    \label{fig:evo_diagram}
\end{figure*}

\section{Evolution of the Dark Comets}\label{sec:evopath}

In this section, we discuss the evolutionary pathway of the dark comets suggested by the limits on material strengths calculated in Sec.~\ref{sec:matstr}.

\subsection{Production via Rotational Fragmentation Cascade}\label{subsec:model}

A rotational fragmentation cascade may explain the small sizes {and rapid rotations} of the dark comets. The weak nongravitational accelerations of these objects suggest that they are highly thermally evolved and largely devolatilized or covered with refractory mantles, if they are cometary in origin. As such objects devolatilize, sublimative torques operating on their nuclei would produce spinup, potentially to their rotational disruption limits \citep{Jewitt1997,Steckloff2016a,Hirabayashi2016,Jewitt2021,
Safrit2021,Jewitt2022dest}. Such rotational disruption could spall off pieces of material from the nucleus. So long as these ejected pieces have less than $\sim$20\% of the mass of the original nucleus, the ejected piece will gravitationally escape the parent nucleus \citep{Hirabayashi2016}, and may continue to rotationally disrupt.

This rotational cascade could produce small fragments that reside in or near the strength-dominated regime. In this case, breakup would produce even smaller fragments that are more resistant to further disruption, because smaller objects require even faster rotation to disrupt (see Eq.~\ref{eq:sigmatdef}). However, continual outgassing torque would continue to spin fragments up to their disruption limits. Eventually, these objects would be sufficiently small and devolatilized to remain stable against further disruption. As a result, one would expect rotationally-disrupted objects to continually spin up until all that remains are small fragments that are stable to further disruption. This final state is consistent with the observed properties of dark comets, which are stable against rotational disruption if they have typical cometary bulk tensile strengths. Therefore, dark comets may represent one of the final endpoints of comet nucleus evolution --- the small ($s\sim1-10$ m), largely devolatilized remnants of their parent nuclei. 

This implies a possible evolutionary sequence for dark comets --- they begin as part of a volatile-rich body that fragments due to rotational disruption. The resulting fragments may then undergo a rotational spin-up cascade until they reach small sizes that are stable against disruption, where they remain until they become devolatilized. At some point during this evolutionary sequence, these objects would {dynamically} transfer from their {original} environment to an orbit with a semimajor axis $a\simeq 1$ au. We show this hypothetical evolutionary sequence in Fig.~\ref{fig:evo_diagram}. 

Comet 252P/LINEAR may be undergoing the first steps toward evolving into a dark comet. 252P is thought to be a fragment of comet 460P/2016 BA$_{14}$ Pan-STARRS, due to their similar orbits \citep{Li2017}. Their relative sizes \citep[300 m for 252P and 1 km for 460P;][]{Li2017} are consistent with this origin, according to the conditions found by \citet{Hirabayashi2016}. 252P is also a fast rotator for a comet nucleus, with a light curve--derived rotation period of only 5.41 h \citep{Li2017}. However, it is worth noting that 460P may have a rotation period as slow as 40 h \citep{Naidu2016}. Because the size ratio of 252P and 460P is $\sim1/3$, 252P may carry away enough angular momentum to slow the rotation of 460P to this rate. Therefore, this slow rotation rate is compatible with this fragmentation mechanism. 

As the sizes of the fragments decrease, they may continue to spin up. Some of these fragments may undergo a sublimation-driven rotational cascade that disrupts them into fine debris, potentially producing dust trails that resemble isolated striae \citep{Steckloff2016a}. Alternatively, these fragments may mostly devolatilize prior to the completion of the rotational fragmentation cascade. In such cases, the fragments would reach small sizes with fast rotation periods and minimal outgassing --- the dark comets.\footnote{{Note that the small number of measured rotation rates on the dark comets makes any comparison highly uncertain.}}

This suggests that the near-Earth space may be littered with numerous dead (fully devolatilized) comet fragments.  Nevertheless, they may be exceedingly difficult to detect, or identify if already detected. In addition to being small and correspondingly dim, the weak sublimation-driven nongravitational accelerations that are the currently-identified hallmark of dark comets \citep{Farnocchia2023,Seligman2023} will fade until their nuclei are completely devolatilized. Thereafter, dark comets would appear only as small, inactive asteroids/meteoroids (``dead comets''). The detected dark comets may therefore represent objects finishing their transition into this ultimate end state. 

This is a significantly different evolutionary path than that taken by larger comets that are predominantly held together by self-gravity \citep[comet nuclei $\gtrsim500$ m in radius, based on][]{Safrit2021}.\footnote{\citet{Bottke2023} found that the strength-gravity transition was at $R\sim20$ m, using a model of collisional evolution of Jupiter Trojans and outer solar system populations, combined with laboratory data. This value is consistent with our results for $\sigma_t\simeq0.1$ Pa (see Fig.~\ref{fig:size-period}).} Such large comet nuclei predominantly evolve via mass-wasting events, which flatten the {surface of the} nucleus over time, expose buried volatiles, and maintain sublimative activity \citep{Vincent2015,Steckloff2018}. As these nuclei flatten, the frequency of outbursts decreases, leading to episodic activity that can reactivate comet nuclei \citep{Steckloff2018}. Eventually, these nuclei can become sufficiently gravitationally flat to preclude further mass-wasting reactivation, and lapse into large dead comets \citep{Steckloff2018}. In addition, the weak relative tensile strength of comets at gravity-dominated scales causes {comets} to be destroyed in the near-Earth environment, preventing gravity-dominated objects from remaining as NEOs {for long periods of time, since close encounters will tend to destroy them} \citep{Levison1997, Fernandez2002, DiSisto2009, Granvik2023, Nesvorny2024a}. 

However, throughout this process, gravity-dominated nuclei generally do not change size. Sublimative spin-up would tend to deform the nucleus \citep{Safrit2021} or fission off gravitationally bound fragments that can reaccrete \citep{Hirabayashi2016}. This is fundamentally different from the small strength-bound dark comets, which can more easily spin up and {completely divide}. Although both of these tracks end with a largely devolatilized nucleus, their sizes and final spin rates are likely to diverge based on the track followed. Under this hypothesis, the evolutionary path of a comet is {strongly dependent on} the size of its nucleus {(and its dynamical history)}, which is consistent with the JFCs, {MBCs}, and dark comets shown in Fig.~\ref{fig:size-period}.

\subsection{Devolatilization and Fission Timescales}\label{subsec:devoltime}

The timescale over which these objects would devolatilize $\tau_{\rm dev}$ is set, at minimum, by the time required for a thermal heat pulse to reach the centers of these objects. \citet{Safrit2021} {found that}
\begin{equation}\label{eq:taudev}
\tau_{\rm dev} = \frac{R^2}{\alpha}\,.
\end{equation}
In Eq.~\eqref{eq:taudev}, $R$ is the radius of the dark comet nucleus and $\alpha$ is the thermal diffusivity of the material, typically on the order of $\alpha\sim 10^{-8}-10^{-7}$ m$^2$ s$^{-1}$ \citep{Gundlach2012, Jewitt2017, Groussin2019, Steckloff2021}. Thus, for dark comets with radii between 2 and 15 m, $\tau_{\rm dev}\sim 1-1000$ years.

We can compare this to the timescale required for sublimative torques to spin up an object, $\tau_{\rm tor}$. Using the sublimative YORP (SYORP) formalism from \citet{Steckloff2016a}, we {write} that
\begin{equation}\label{eq:tautor}
    \tau_{\rm tor} = \frac{R\sqrt{32\rho \sigma_t}}{P_S Y_S}\,.
\end{equation}
In Eq.~\eqref{eq:tautor}, $P_S$ is the material sublimation pressure \citep{Steckloff2015, Steckloff2016a} and $Y_S$ is the SYORP parameter, which measures the fraction of the sublimative momentum flux that must be directed tangentially to produce an equivalent torque. In the near-Earth environment, the sublimation pressure $P_S$ for water ice is on the order of $P_S\sim 0.01$ -- $0.1$ Pa \citep{Steckloff2016a}. SYORP parameters for objects of this size are thought to be on the order of $Y_S\sim 10^{-4}$ -- $10^{-3}$ \citep{Steckloff2016a}, but have been measured in larger, strength-dominated comet nuclei to be an order of magnitude lower, at $Y_S\sim 10^{-5}$ -- $10^{-4}$ \citep{Steckloff2021}. This results in SYORP timescales (time required to monotonically spin up from rest to disruption) of $\sim 2-200$ years for the dark comets with known rotation periods. However, this may be up to an order of magnitude longer if SYORP parameters are weaker. Furthermore, the stochastic nature of torque directions on comets over secular timescales \citep{Hirabayashi2016} may lengthen this by an additional order of magnitude. 

Regardless, this spin-up timescale is comparable to the devolatilization timescale. 
By dividing Eqs.~\eqref{eq:taudev} and \eqref{eq:tautor}, and plugging in typical values, we find that
\begin{equation}\label{eq:tauscale}
\begin{split}
    \frac{\tau_{\rm dev}}{\tau_{\rm tor}}=1.25&\Big(\frac{R}{5\text{ m}}\Big)\Big(\frac{P_S}{0.1\text{ Pa}}\Big)\Big(\frac{Y_S}{10^{-4}}\Big)\Big(\frac{\alpha}{10^{-7}\text{ m$^2$/s}}\Big)^{-1}\\
    &\times\Big(\frac{\rho}{500\text{ kg/m$^{3}$}}\Big)^{-1/2}\Big(\frac{\sigma_t}{10\text{ Pa}}\Big)^{-1/2}\,.
\end{split}
\end{equation}
It is then plausible that while many of these objects can experience rotational disintegration, many others will devolatilize prior to completing a fragmentation cascade, as suggested by \citet{Steckloff2016a}. In this case, the cascade would terminate with a small, rapidly-rotating dead comet --- a dark comet without any nongravitational acceleration. 

\begin{figure}
    \centering
    \includegraphics{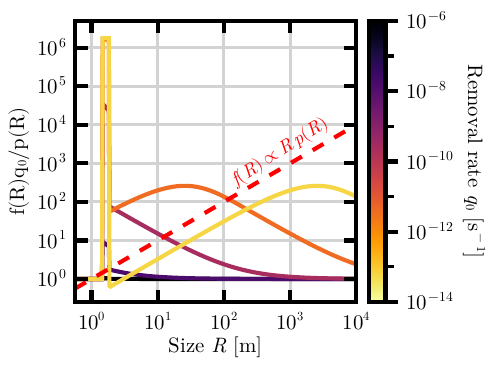}
    \caption{The normalized size frequency distribution (SFD) of the objects evolving under this model, for a range of removal rate parameters $q_0$. We show $q_0\in[10^{-14}, 10^{-6}]$ s$^{-1}$. We expect $q_0=10^{-14}$ s$^{-1}$ to be the true removal rate, and $q_0=10^{-6}$ s$^{-1}$ is chosen because the distribution is flat at this point. The model SFD $f(R)$ is shown in terms of the steady-state divisionless SFD, $p(R)/q_0$. We assume that the injection SFD power-law slope is $\beta=1$, but this only affects the exact values rather than the slope. The predicted small-R limit of $f(R)\propto R\, p(R)$ is shown as a red dashed line. }
    \label{fig:SFD}
\end{figure}

\section{Size Frequency Distribution}\label{sec:sfd}

In this section, we estimate a size frequency distribution (SFD) for objects undergoing the rotational fragmentation cascade {in the NEO environment} discussed in Sec.~\ref{sec:evopath}. 

We consider an idealized case where the rotational cascade only splits objects in half. Specifically, each splitting event replaces an object of radius $R$ with two objects of radius $R/(2^{1/3})$. We ignore collisional fragmentation in these calculations because the collisional timescale \citep{Bottke2005} is 
\begin{equation}
    \tau_{\rm col}= 1\text{ Myr }\left(\frac{R}{10\text{ m}}\right)^{1/2}\,.
\end{equation}
Since $\tau_{\rm div}\ll \tau_{\rm col}$ for the relevant sizes, collisions are negligible. {We then define the} differential object fraction per radius $R$ {as}
\begin{equation}
    f(R)\equiv\frac{\de n}{\de R}\,.
\end{equation}
This is normalized such that the total number of objects $N=\int_0^\infty\!\! f(R)\,\de R$. 

Next, we define a function $p(R)$, which is the number of objects of a size $R$ that are injected into the population over a time step $\de t$. This function is set by the input from the {source population} into the NEO environment. {Note that this analysis is agnostic to the source population and is applicable to a JFC or MBC source.} We further define a constant $q_0$, which is the fraction of objects that are removed from the population in a given time $\de t$ by orbital instabilities. This is assumed to be constant across all object sizes.

Finally, the population of objects of a given size is determined by the rotational cascade. We define a function $q(R)$, which is the fraction of objects of size $R$ that divide in a time step $\de t$. Combining these definitions, we find that the change in the population is given by 
\begin{equation}\label{eq:dfdt}
    \frac{\de f}{\de t}(R)=p(R)+2q(\epsilon R)f(\epsilon R)-q(R)f(R)-q_0 f(R)\,.
\end{equation}
The first term {on the right-hand side} is the number of objects injected {into the population}, the second term is the addition of two objects via splitting of an object with size $\epsilon R$, the third term is the removal of objects via fragmentation events, and the final term is the removal of objects via orbital instabilities. In this case, $\epsilon=2^{1/3}$. In order to find the steady-state SFD, we set $\de f/\de t=0$.

We assume that $p(R)$ is constant in time, so that the age distribution of objects is uniform. Therefore, if objects live in the NEO environment for $10^6$ yr \citep{Gladman2000, Nesvorny2023}, then $q_0\simeq 10^{-14}$ s$^{-1}$. 

There are two competing factors to determine $q(R)$: spin-up and devolatilization.  The timescales for these processes are generally comparable (Sec.~\ref{subsec:devoltime}, {Eq. \eqref{eq:tauscale}}), but fragmentation does not occur when the devolatilization timescale is shorter than the spinup timescale. However, if the spinup timescale is shorter than the devolatilization timescale, then $f(R)\de t/\tau_{\rm div}(R)$ objects are removed, where $\tau_{\rm div}$ is the spinup timescale. We assume that {all} objects begin at the critical rotation rate of the progenitor object, even if they {were} recently inserted {into the population}. Therefore, $\tau_{\rm div}=\tau_{\rm tor}(1-1/\epsilon)$, where $\tau_{\rm tor}$ is given by Eq.~\eqref{eq:tautor}, and the factor of $(1-2^{-1/3})$ is introduced by beginning at $\omega=\omega_{\rm crit}(\epsilon R)$ rather than $\omega=0$. With the devolatilization timescale given by Eq.~\eqref{eq:taudev}, $q(R)$ is given by
\begin{equation}\label{eq:qRdef}
    q(R)=\begin{dcases}
        \frac{P_S Y_S}{\sqrt{32\rho\sigma_t}}\frac{1}{R}\frac{1}{(1-1/\epsilon)} & \tau_{\rm div}\leq \tau_{\rm dev}\\
        0 & \text{otherwise}\,.
    \end{dcases}
\end{equation}

Solving Eq.~\eqref{eq:dfdt} for $f(R)$ is complicated by the fact that both $f(R)$ and $f(\epsilon R)$ appear in the equation. Moreover, a single solution does not exist because $q(R)$ is a piecewise function of radius $R$. We therefore consider three domains of the function. 

\textbf{Case 1.} First we consider the case where $R$ is small and $q(R)=q(\epsilon R)=0$. In this instance, {Eq. \eqref{eq:dfdt} can be easily solved to show that}
\begin{equation}\label{eq:fI}
    f(R) = \frac{p(R)}{q_0}\,.
\end{equation}
In this limit, $f(R)$ has identical behavior to the input function, albeit modified by a factor that accounts for the constant removal {by orbital instability.}

\textbf{Case 2.} Next, we consider the case where $R$ is large and $q(R)\neq0$, at the extreme end from the previous case. We can rewrite Eq.~\eqref{eq:dfdt}, solving for $f(r)$ in terms of $f(\epsilon R)$ to obtain 
\begin{equation}\label{eq:fII}
    f(R)=\frac{p(R)+2q(\epsilon R)f(\epsilon R)}{q_0+q(R)}\,.
\end{equation}
We continue this series, writing $f(\epsilon R)$ in terms of $f(\epsilon^2 R)$ up to infinity. From Eq.~(\ref{eq:qRdef}), $q(R)=A/R$ where $A=(1-2^{-1/3})^{-1} P_S Y_S/\sqrt{32\rho\sigma_t}$ is a constant. Therefore, $q(\epsilon^k R)=\epsilon^{-k} q(R)$. We also assume that $p(R)$ is a power law such that $p(R)\propto R^{-\beta}$. Continuing out the series and simplifying, we find that 
\begin{equation}\label{eq:fIII}
\begin{split}
    f(R)=p(R)\sum_{n=0}^{\infty}&2^n \epsilon^{-n(2\beta+n+1)/2} q(R)^n\\
    \times&\prod_{k=0}^n \Big[q_0+q(\epsilon^k R)\Big]^{-1}\,.
\end{split}
\end{equation}
In general, Eq.~\eqref{eq:fIII} can be used to compute $f(R)$ for the domain where $q(R)\neq0$, so long as the series converges.

\textbf{Case 3.} In the final case, $q(R)=0$ and $q(\epsilon R)\neq0$. This domain is transitional between the other two domains considered, and occurs for middling values of $R$. In this instance, Eq.~\eqref{eq:fII} can be used to refer $f(R)$ to $f(\epsilon R)$. By construction, $f(\epsilon R)$ falls into the large-$R$ regime and can be solved by evaluating Eq.~\eqref{eq:fIII}. 

{We show $f(R)$ for a range of $q_0$ values in Fig. \ref{fig:SFD}. We normalize the SFD to the small-$R$ limit $p(R)/q_0$. We assume that $\beta=1$, $P_S=0.1$ Pa, $Y_S=10^{-4}$, $\sigma_t=50$ Pa, $\rho=500$ kg/m$^{3}$, $\alpha=10^{-7}$ m$^2$ s$^{-1}$, and $\epsilon=2^{1/3}$, identically to Eq.~\eqref{eq:tauscale}.} For small $R$, the {form} of the SFD is unchanged from the input function (Eq.~\ref{eq:fI}). For large $R$, the relationship between Eq.~\eqref{eq:fIII} and $p(R)$ depends on the removal parameter $q_0$. If $q_0=0$, then  $f(R)\propto R p(R)$ and the power-law index is increased by a factor of order unity. Similarly, if $q_0\neq0$ then $f(R)\propto p(R)$. As $q_0\rightarrow0$, the small-$n$ terms will be dominated by $R p(R)$, before eventually converging to $p(R)$ at large $R$. Therefore, the value of $q_0$ determines the slope of the power law for large objects. There are significantly more objects undergoing evolution than are injected, due to the relatively low values of the removal rate $q_0$. The spikes at $R\sim 1$ m correspond to the cutoff where objects {are too small to be} removed by division. At even smaller sizes there is another cutoff as the population is no longer fed by the division of larger objects. 

\begin{table*}[t]
\caption{\textbf{Dark Comet Source Probabilities.} {The probability that each dark comet originates from a given source based on our dynamical models. These results should be considered order of magnitude estimates, since (i) the source populations are normalized to the NEOMOD dataset and are not adjusted to our model's modified orbital distributions and (ii) these probabilities do not account for the composition of the dark comets and the source populations, which may significantly modify estimations of their true origins. These source probabilities are generally consistent with the NEOMOD model \citep{Nesvorny2024b} and the model of \citet{Granvik2018}.}} 
\label{table:dcprobs}
\begin{tabular}{lrrrrrrrrrrr }
 Object & JFCs & $\nu_6$ & 3:1 & 5:2 & 7:3 & 8:3 & 9:4 & 11:5 & 2:1 & Forest & High $i$ \\
 \hline
1998 KY$_{26}$ & 0 & 0.893 & 0.088 & 0.002 & 0 & 0.001 & 0 & 0.001 & 0 & 0.012 & 0.004 \\
2005 VL$_1$ & 0 & 0.850 & 0.125 & 0.010 & 0 & 0.001 & 0 & 0.001 & 0 & 0.012 & 0.001 \\
2016 NJ$_{33}$ & 0 & 0.831 & 0.096 & 0.049 & 0 & 0.002 & 0 & 0 & 0 & 0.020 & 0.003 \\
2010 VL$_{65}$ & 0 & 0.797 & 0.198 & 0 & 0 & 0 & 0 & 0 & 0 & 0.004 & 0.001 \\ 
2010 RF$_{12}$ & 0 & 0.948 & 0.026 & 0.010 & 0 & 0.002 & 0 & 0 & 0 & 0.010 & 0.004 \\ 
2006 RH$_{120}$& 0 & 0.389 & 0.505 & 0 & 0 & 0 & 0 & 0 & 0 & 0.106 & 0 \\
2003 RM & 0.011 & 0.001 & 0.040 & 0.911 & 0.005 & 0.010 & 0.003 & 0.019 & 0 & 0 & 0 \\\hline
\end{tabular}
\end{table*}

\begin{figure*}[t]
    \centering
    \includegraphics{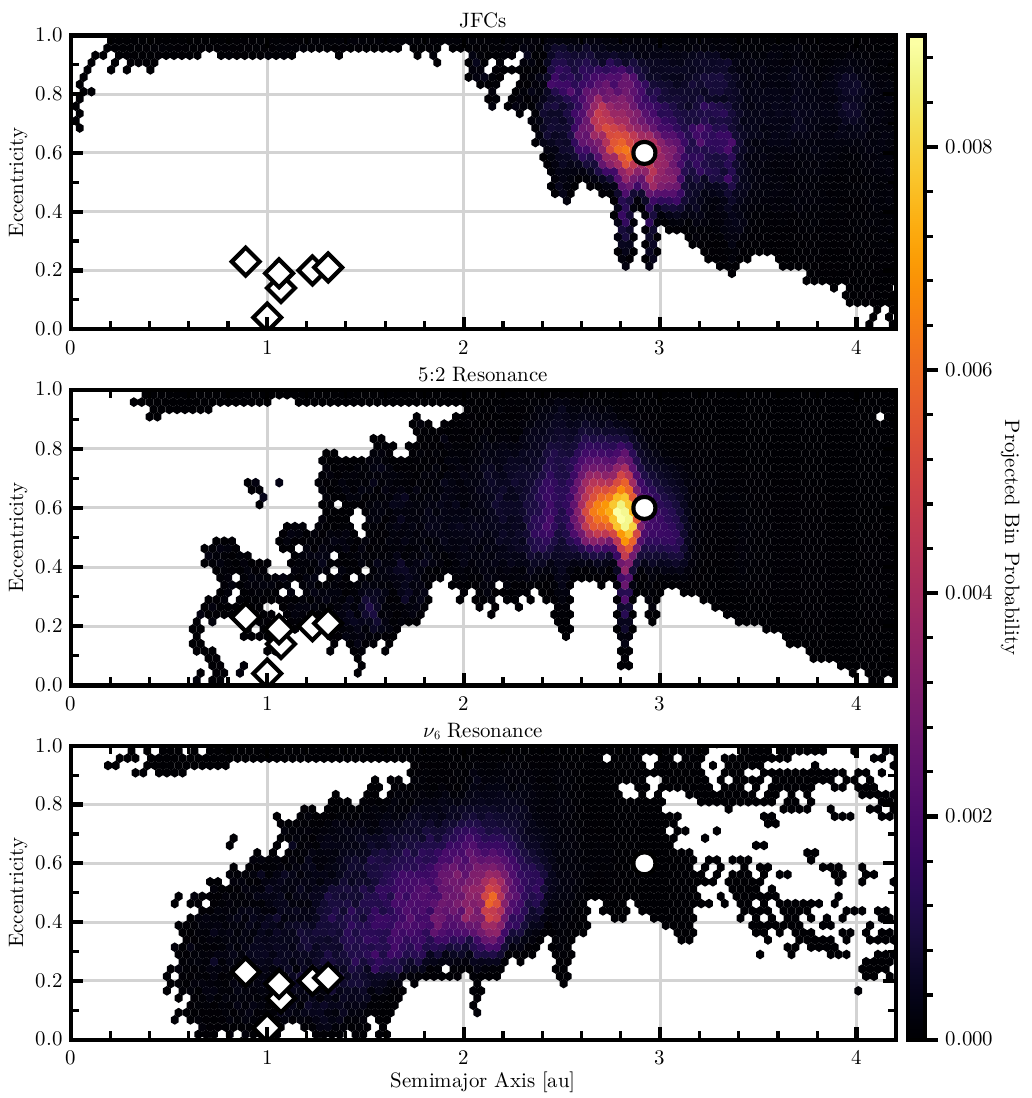}
    \caption{{The bin probabilities for JFCs and main belt objects {(from the $\nu_6$ and 5:2 resonances)} entering the near-Earth environment. The populations are initialized from the NEOMOD model \citep{Nesvorny2023} and then integrated further with nongravitational accelerations. A white background indicates that no object reached that point in the simulations. Only samples with an inclination between $0^\circ$ and $15^\circ$ are shown. The large white circle is 2003 RM, which has an orbit consistent with a JFC {or 5:2 resonance} origin. The white diamonds are the other, shorter-period dark comets, {which are the most similar to a $\nu_6$ origin}.}}
    \vspace{-15pt}
    \label{fig:sourcesims}
\end{figure*}

\section{Dynamical Origins}\label{sec:dynorig}

In this section, we use the NEOMOD model \citep{Nesvorny2023} to investigate the dynamical origins of the dark comets. While NEOMOD makes predictions for the source population for a given set of orbital parameters, this model may not be fully accurate for the dark comets. Principally, NEOMOD does not include nongravitational accelerations, which are the defining feature of dark comets.

As a result, we run dynamical simulations with nongravitational accelerations, using initial conditions from the NEOMOD model. We consider 11 of the 12 initial populations in NEOMOD --- the Hungarias and Phocaeas are combined into a single high-inclination source. We therefore use the JFCs, the high-inclination sources, the $\nu_6$ resonance, the 3:1 resonance, the 5:2 resonance, the 7:3 resonance, the 8:3 resonance, the 9:4 resonance, the 11:5 resonance, the 2:1 resonance, and the forest of weak resonances in the inner main belt as individual sources.

For each source, we generate a representative initial sample of $10^4$ objects, which are chosen from the probability density function (PDF) given by the NEOMOD data. We set each object to have a nongravitational acceleration of the form described in Appendix \ref{app:nongravform} with a magnitude of $|A_0|=10^{-5}$ au yr$^{-2}$, consistent with the nongravitational accelerations of the dark comets (see Table \ref{table:objects}). Each source population is then numerically integrated for $10^5$ yr, consistent with the mass budget described in Appendix \ref{app:massbudget}. Models are constructed for each source following the methodology described in Appendix \ref{app:simdetails}.

From this dataset, we calculate a PDF for comparison with the dark comets. We then normalize each source population by multiplying each source's PDF by the fraction of all NEOs in that source (according to NEOMOD). This then provides the probability for each dark comet to originate from a given source. A more detailed discussion of this methodology is given in Appendix \ref{app:probcalcs}. These probabilities are given in Table \ref{table:dcprobs}, although they are limited to order-of-magnitude estimates by several factors. First, our methodology of renormalization may not be accurate, since our population samples are not identical to those implemented in NEOMOD. Second, these probabilities do not account for the different degrees of volatile enrichment in these populations {but} only reflect our dynamical calculations. Since the dark comets are most likely outgassing, the volatile enrichment of these populations is an important factor to consider in identifying their origins {and} interpretation of these probabilities must be done with care. Future observations of the dark comets' surface features and outgassing properties (species and rates) will constrain the source populations of the dark comets beyond the dynamics alone.  

{In Fig. \ref{fig:sourcesims}, we show the PDF\footnote{{Note that this is not the probability that an object is from a source given its bin, but is the probability of an object being found in the bin given its source. The difference is subtle, but important. See Appendix \ref{app:probcalcs} for more details. }} for various source populations as a function of semimajor axis and eccentricity, with the dark comet locations indicated. Specifically, we show {PDFs} for the JFCS, the $\nu_6$ resonance, and the 5:2 resonance (see Appendix \ref{app:simdetails} for more complete source region calculations).} We {restrict the PDF to an} inclination of $i\in[0, 15]^\circ$, which includes all of the dark comets (see Table \ref{table:objects}).\footnote{Note that the analogous calculations in Appendix \ref{app:simdetails} do not include an inclination cutoff.} 2003 RM is {most likely a main belt object,} specifically from the nearby 5:2 resonance, {but is also located where a JFC is most likely to occur}. {At this time, it appears that 2003 RM is volatile enriched and could plausibly originate in either the JFC region or the main belt, {presumably via an MBC}. Follow-up observations may help to constrain the origin of this object.} 

The remaining dark comet orbits are consistent with a main belt source, particularly the $\nu_6$ source in the inner main belt. However, the retention of volatiles in the inner main belt is more challenging, {suggesting the presence of undiscovered MBCs or subsurface volatiles on asteroids in this region}. If the dark comets are  produced by a rotational fragmentation cascade, then a single object en route to a year-period orbit may produce several orders of magnitude more {small objects}, increasing the prevalence of this population {and accounting for the number of dark comets}.

These simulations show a significant mixing of the NEO source populations when nongravitational accelerations are incorporated. This makes it challenging to identify a dynamical origin based on orbital elements alone. Other characteristics --- color, albedo, nongravitational accelerations, material strength, etc. --- will be necessary to clarify any given object's dynamical origin. Given these results, the inner dark comets likely originated from the main belt. {While 2003 RM's dynamics imply an origin in the outer main belt, a JFC origin is also plausible.} Given the apparent degeneracies in source population probabilities, follow-up observations of {physical} properties may aid in the identification of the dark comets' source populations. 

\begin{figure}[t]
    \centering
    \includegraphics{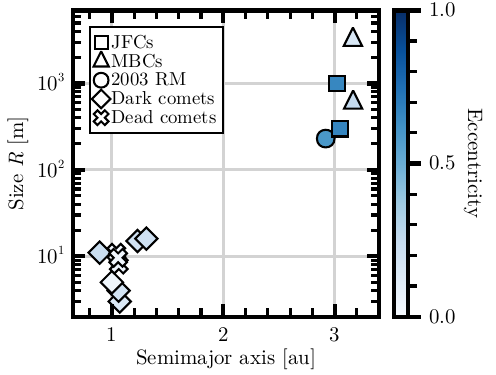}
    \caption{Sizes and semimajor axes of dark comets, the JFCs 252P/LINEAR and 460P/PANSTARRS, {the MBCs 133P/Elst-Pizarro and 238P/Read}, and the possible dead comets. The color indicates the eccentricity. }
    \label{fig:allobjs}
\end{figure}

\section{Discussion}\label{sec:discussion}

In this paper, we investigate the physical and dynamical origins of the ``dark comets'' identified by \citet{Farnocchia2023} and \citet{Seligman2023}. We propose that {these objects} begin as smaller, strength-dominated {objects} and evolve through gravitational effects into the NEO environment. During this transition, and especially once in the NEO environment, these progenitor objects undergo significant nongravitational accelerations {that} cause a rotational fragmentation cascade. The resulting fragments would continue to spin up and devolatilize. However, the rotational torque is reduced as the volatiles are depleted. Eventually, the torque will be insufficient to spin the object up beyond its critical rate and it will be stable against further rotational destruction. At this point, the fragments will be small, rapidly rotating, and mostly devolatilized. 

{While small-number statistics make comparison difficult,} there are several lines of observational evidence {that suggest that the dark comets may be the product of a rotational fragmentation cascade.} {These objects have small radii, weak nongravitational accelerations, and  rapid rotation rates (where measured). Notably, these rotation periods are above the stability threshold for larger objects. Therefore, these objects may have been generated by the fragmentation of a larger object.} We also calculate the SFD of the dark comets under the action of the rotational fragmentation cascade. The SFD has two observable signatures: (i) a modification to the {power-law} slope of the injected objects' {SFD} and (ii) a sharp increase in the number of objects with sizes of $R\sim 1$ -- $10$ m. The second of these {signatures} is consistent with the currently known population of dark comets, although limited by small-number statistics. 

This model may also explain the origins of several NEOs with low densities and detected nongravitational acceleration due to radiation pressure --- 2009 BD, 2012 LA, and 2011 MD. These objects are small ($R\sim10$ m) and have densities of $\rho\simeq(500\pm300)$ kg/m$^{3}$, consistent with comets \citep{Richardson2007, Micheli2012, Micheli2013, Micheli2013thesis, Micheli2014,Patzold2016}. {\citet{Mommert2014a} and \citet{Mommert2014b} found larger densities consistent with asteroids for 2009 BD and 2011 MD, although 2011 MD's density is still consistent with 500 kg/m$^{3}$.} The extinct devolatilized comet fragments predicted in this model should be meter-scale with {relatively} low densities, consistent with these {NEOs}. These bodies, as well as the possible precursor JFCs 252P/LINEAR and 460P/PANSTARRS and {volatile-active} MBCs 133P/Elst-Pizarro and 238P/Read, are shown in Fig.~\ref{fig:allobjs} along with the dark comets. 

{The rotational fragmentation model presented here both is  self-consistent and explains the observed properties of the dark comets. However, there are significant sources of uncertainties in the current data and rotation periods are only measured on a few dark comets. As a result, alternative models may be similarly consistent with the limited data.} {For example,} tidal disruption may also {be responsible for the creation of the} dark comets. A larger rubble-pile comet could undergo tidal disruption through a close encounter with a terrestrial planet, producing a family of smaller NEOs with similar sizes to the dark comets \citep{Granvik2023, Nesvorny2024a}. These events will produce an overdensity of meter-scale objects, similar to our predicted rotational-fragmentation SFD (Sec.~\ref{sec:sfd}). Both tidal and rotational disruption may therefore be responsible for producing dark comets, depending on the relative size and binding force of the progenitor {object}.

{The rotational fragmentation evolutionary} track predicts a relatively large population of devolatilized {fragments} in the NEO environment, either with undetectably low nongravitational accelerations (dark comets) or fully extinct (dead comets). {However,} this calculation relies on the residence time and mass-loss time of these objects, which must be adjusted for the presence of nongravitational accelerations. Moreover, our estimates only apply to small NEOs that are capable of being produced by this model (Sec.~\ref{sec:sfd}). While this {prediction} is consistent with our estimate that 0.5 -- 60\% of NEOs could originate along this pathway, more work is needed to refine this value.

{If these objects are generated by a rotational fragmentation cascade,} few objects should be observed in the interim stage between the large, relatively young progenitors and the rapidly-rotating, small, thermally evolved inner dark comets. The rotational cascade timescale is $\sim10^3$ yr, significantly shorter than the residence time in the near-Earth environment of $\sim10^6$ yr \citep{Gladman2000, Nesvorny2023}. There should therefore be two distinct populations of dark comets --- the progenitors and the fragments. The progenitors are relatively large (100 -- 1000 m), new to the near-Earth environment, and may have only recently begun to sublimate, not yet triggering the rotational fragmentation cascade. Meanwhile, the small fragments have likely finished this cascade and are mostly devolatilized. Since these evolutionary stages have long lifetimes compared to the rotational {fragmentation} cascade, most of the dark comets will be observed in these two edge states. 

{This model also explains the outlier of 2003 RM, which is larger and} has a {primarily} transverse {nongravitational} acceleration. {In contrast,} the smaller dark comets primarily have out-of-plane {nongravitational} components. This feature (although limited by small-number statistics) can be understood in the context of this model. Volatile-rich objects will likely have radial and transverse nongravitational accelerations when they first {become active}. These accelerations may spin up the objects and trigger a rotational fragmentation cascade. When the nuclei are rotating sufficiently rapidly to have longitudinally isothermal surfaces, the outgassing can align with the spin axis and produce out-of-plane accelerations \citep{Taylor2024}. The {size of 2003 RM and the} failure of the outgassing balancing mechanism to reproduce {its} acceleration {are} therefore consistent with this model, implying that 2003 RM is in the beginning stage of this evolutionary track and has not been spun-up to the isothermal limit.

We also investigate the dynamical origins of these objects. The orbit of 2003 RM implies that {it likely originated in the outer main belt (see Table \ref{table:dcprobs}), although there are several caveats to this conclusion. First, the orbit of 2003 RM resides precisely at the maximum probability for a JFC origin (see Fig. \ref{fig:sourcesims}). Second, the relatively low probability of a JFC origin compared to a main belt origin  primarily reflects the small number of JFCs relative to main belt objects (Table \ref{table:PSjs}). Finally, if the nongravitational acceleration of 2003 RM is driven by outgassing of volatiles, then the volatile-rich JFCs are a more likely a priori source. As a result, the JFCs are still a plausible source for 2003 RM.} 

Meanwhile, our numerical experiments demonstrate that the smaller {dark comets} likely originated in the {inner} main belt. {This may have significant ramifications concerning the existence and abundance of volatiles in this region. While it is already understood that subsurface volatiles can exist in the inner main belt \citep{Fanale1989, Schorghofer2008} and survive into the near-Earth environment \citep{Schorghofer2020}, confirmation of this volatile reservoir remains elusive. The dark comets potentially provide evidence that a volatile reservoir exists in the inner main belt. Furthermore, the presence of these objects in the near-Earth environment potentially provides an additional pathway for terrestrial volatile delivery.}

Ground- and space-based follow-up observations of dark comets may enable measurement of the outgassing rates and compositions, potentially constraining their dynamical origins. Spectral observations {in particular} may provide more information regarding their progenitor asteroid spectral type {and the size of the main belt volatile reservoir.} Future survey missions such as \textit{NEO Surveyor},  the Rubin Observatory Legacy Survey of Space and Time (LSST), and the \textit{Gaia} mission may also identify more dark comets than are currently known, which will refine our understanding of this evolutionary pathway {and their source populations}. 

Finally, the Hayabusa2 Extended Mission (Hayabusa2\#) will arrive at 1998 KY$_{26}$ in 2031 \citep{Hirabayashi2021, Kikuchi2023}. If the {rotational fragmentation} model is correct, then Hayabusa2\# should find an object with high porosity and a material strength of $\sigma_t\geq 5.3$ Pa, similar to {a cometary body}. In addition, the rotation period of 1998 KY$_{26}$ was last measured in 1999 \citep{Ostro1999}. As a result, the SYORP effect may have further spun up 1998 KY$_{26}$ in the intervening time, which would be detectable by Hayabusa2\# and further radar observations. This will provide another constraint on the accuracy of {the rotational fragmentation} model. {Given the dynamical origins and nongravitational acceleration of 1998 KY$_{26}$, this mission will also provide an opportunity to further investigate and characterize the unique volatile reservoir in the inner main belt.} 

\section*{Acknowledgements}

We thank the two anonymous reviewers for their helpful suggestions, which significantly improved the scientific content of this manuscript. {We also thank the editor of this manuscript, Lori Feaga, for her assistance in the review process.} We thank Fred Adams, Juliette Becker, Bill Chen, Fei Dai, Adina Feinstein, Thomas Kennedy, Garrett Levine, Nikole Lewis, Kevin Napier, Luis Salazar Manzano, Cindy Xiang, and Andrew Youdin for useful conversations and suggestions. A.G.T. thanks Fiona Corcoran for assistance with figures. 
A.G.T. acknowledges support from the Fannie and John Hertz Foundation and the University of Michigan's Rackham Merit Fellowship Program. J.K.S. acknowledges support from NASA Grant No.\ 80NSSC19K1313 and NASA Grant No.\ 80NSSC22K1399. D.Z.S. is supported by an NSF Astronomy and Astrophysics Postdoctoral Fellowship under award AST-2202135. This research award is partially funded by a generous gift of Charles Simonyi to the NSF Division of Astronomical Sciences. The award is made in recognition of significant contributions to Rubin Observatory’s Legacy Survey of Space and Time. D.F. conducted this research at the Jet Propulsion Laboratory, California Institute of Technology, under a contract with the National Aeronautics and Space Administration (80NM0018D0004). L.D. acknowledges support from an internal research grant at Southwest Research Institute. 

{This paper made use of the \texttt{Julia} programming language \citep{Julia} and the plotting package \texttt{Makie} \citep{Makie}. Simulations in this paper made use of the \texttt{REBOUND} N-body code \citep{rebound} and the \texttt{REBOUNDx} library \citep{reboundx}. The simulations were integrated using \texttt{MERCURIUS} \citep{mercurius}.}

\bibliographystyle{aasjournal.bst}
\bibliography{main}

\appendix
\restartappendixnumbering

\section{Stochastic Nongravitational Acceleration}\label{app:nongravform}

{In this section, we introduce the model of nongravitational acceleration that we use in our dynamical simulations. Our model is loosely based on the model of \citet{Fernandez2002}, who showed that nongravitational accelerations may allow objects to decouple from Jupiter and migrate inwards. We assume that the nongravitational acceleration takes the form }
\begin{equation}
   {\boldsymbol{F}=\alpha A_0\left(\frac{r_0}{r}\right)^2\boldsymbol{\hat{e}}_v}\,. \label{eq:nongravform}
\end{equation}
In Eq. \eqref{eq:nongravform}, $\boldsymbol{\hat{e}}_v$ is the unit velocity vector, $r_0$ is a scaling distance (typically 1 au), $r$ is the heliocentric distance, and $\alpha$ is a constant between -1 and 1. The parameter $\alpha$ is introduced in order to account for stochastic variation in the direction and magnitude of the nongravitational acceleration. Here, this parameter is randomly chosen from a sinusoidal distribution and resampled every of 50 yr. We have implemented this nongravitational acceleration in \texttt{REBOUNDx} \citep{reboundx}.

\section{Outgassing Mass Budget}\label{app:massbudget}

{In this section, we calculate a mass budget for the known dark comets, which we convert into an allowable nongravitational acceleration magnitude over a timescale.}

The mass-loss rate by outgassing of a species $X$ is given by \citep{Seligman2023}
\begin{equation}
    {\frac{\de M}{\de t}=\frac{M |A_i|}{v_{\rm gas}\zeta}}\,. \label{eq:Mdot}
\end{equation}
{The variable $\zeta$ indicates the collimation of the outgassing, and $v_{\rm gas}$ is the gas velocity, given by }
\begin{equation}
    {v_{\rm gas}=\left(\frac{8k_B T_{\rm gas}}{\pi m_X}\right)^{1/2}}\,. \label{eq:vgas}
\end{equation}
{Assuming that $A_i$ and $v_{\rm gas}$ are constants in time, Eq.~\eqref{eq:Mdot} can be solved to find that }
\begin{equation}
    {M(t)=M_0\exp\left(-\frac{|A_i|t}{v_{\rm gas}\zeta}\right)}\,. \label{eq:Msol}
\end{equation}
Technically, there is no true time limit for a nongravitational acceleration with a fixed magnitude --- as the object's mass decreases, the acceleration requires less mass outflow, and so $M\rightarrow0$ only when $t\rightarrow\infty$. However, we are here restricted by our (assumed) knowledge of the {initial and} final states of the dark comets --- objects of $R_{\rm nuc}\sim 200$ m {become objects of $R_{\rm nuc}\sim 10$ m}. {Assuming that the dark comets are spheres, the acceleration and the timescale are therefore constrained by}
\begin{equation}
    |A_i|\tau\leq 3 v_{\rm gas}\zeta \ln\left(\frac{R_0}{R(\tau)}\right)\,.
\end{equation}
{This restriction implies that we can solve for a relationship between the nongravitational acceleration magnitude $|A_i|$ and the time $\tau$. With $m_X=18$ amu (for H$_2$O outgassing) and $T_{\rm gas}\simeq 100$ K, we find that $v_{\rm gas}\simeq 350$ m s$^{-1}$. Assuming that $\zeta=1$, we can require that}
\begin{equation}
    {|A_i|\,\tau\leq 1\text{ au/yr}}\,.
\end{equation}
{The typical {(nonradiative)} nongravitational acceleration of a dark comet is $|A_i|\simeq10^{-5}$ au yr$^{-2}$, which can therefore be maintained for $\tau\simeq 10^5$ yr. }

It is worth noting that the stochastic outgassing does not perturb an orbit as significantly as a constant nongravitational acceleration in a single direction. The stochastic model is more representative of an ensemble of objects, allowing us to effectively explore the diffusion of these objects in orbital parameter space. The constant acceleration model would instead cause the population to follow {well-defined paths in parameter space}.

\section{Simulation Details}\label{app:simdetails}

In this section, we discuss the details of the numerical simulations that we use to investigate the dark comets' dynamical origins. 

As discussed in Sec. \ref{sec:dynorig}, each simulation is constructed from sources in NEOMOD. {We use 11 of the 12 sources available in NEOMOD --- for convenience, the high-inclination Hungarias and Phocaeas are combined into a single high-inclination source. The other 10 sources --- the JFCs, the forest of weak resonances in the inner main belt, and the $\nu_6$, 3:1, 5:2, 7:3, 8:3, 9:4, 11:5, and 2:1 resonances are included directly.} 

From each source distribution, $10^4$ objects are generated as an initial population. While the NEOMOD sample determines the semimajor axis, eccentricity, and inclination, the other three orbital parameters (longitude of perihelion, longitude of the ascending node, and true anomaly) are chosen uniformly from $[0, 2\pi)$. This initial sample is then loaded into \texttt{REBOUND} \citep{rebound}, along with the Sun and all 8 solar system planets. Using \texttt{REBOUNDx} \citep{reboundx}, each object is given a nongravitational acceleration of the form described in Appendix \ref{app:nongravform} and the typical {(statistically significant, {nonradiative})} magnitude of the dark comets ($10^{-5}$ au yr$^{-2}$; see Table \ref{table:objects}).

{While we do not include radiation pressure or the Yarkovsky effect in these simulations, these forces will not significantly affect our results. These forces are one to three orders of magnitude smaller than the dark comet nongravitational accelerations (see Sec. 2). Second, the larger radiation pressure force only operates in the radial direction and so affects the mean anomaly at epoch of an orbit. For sufficiently small nongravitational accelerations, the radiation pressure will not modify the semimajor axis, eccentricity, or inclination.}

Each source population is then integrated for $10^5$ yr, which is consistent with the mass budget described in Appendix \ref{app:massbudget}. {Our simulations use the \texttt{MERCURIUS} integrator \citep{mercurius}, which uses the \texttt{WHFast} integrator for most circumstances and the \texttt{IAS15} integrator for close approaches.} The sample bodies are set to have no mass and do not interact with each other, although they will interact with the Sun and the planets. If an object collides with a planet, passes within $10^{-2}$ au of the Sun, or passes $>70$ au from the Sun, it is removed from the simulation. The simulation timestep is set to a small value of $48$ h. Every $100$ yr, the semimajor axis, eccentricity, and inclination of each object is recorded, providing a significant sample population {of $N\sim10^7$ points}. {We show the initial and final samples for all of these sources in Figs. \ref{fig:sim1}--\ref{fig:sim11}. In Fig. \ref{fig:sim12}, we show the samples for the main belt as a whole, which was obtained by combining and normalizing all of the samples from the main belt resonances.}

\begin{table}[t]
\caption{\textbf{NEO Source Probabilities.} {The probability that a randomly chosen NEO originates from a certain source population, according to the NEOMOD model. This is equivalent to the fraction of all NEOs that originate in that population. }} 
\label{table:PSjs}
\begin{tabular}{ rl }
 Pop. & Prob. \\
 \hline
 JFCs & 0.0128 \\
 High $i$ & 0.0137 \\
 $\nu_6$ & 0.6003 \\
 3:1 & 0.2979 \\
 5:2 & 0.3570 \\
 7:3 & 0.0012 \\
 8:3 & 0.0082 \\
 9:4 & 0.0008 \\
 11:5 & 0.0072 \\
 2:1 & 0.0032 \\
 Forest & 0.0187 \\
\hline
\end{tabular}
\end{table}

\section{Source Probability Calculations}\label{app:probcalcs}

In this section, we describe our methodology to calculate the source probabilities provided in Table \ref{table:dcprobs}. 

We determine the probability that an object in a parameter-space bin $B_i$ belongs to an arbitrary source population $S_j$. In other words, we calculate the dependent probability $P(S_j|B_i)$. Our simulations provide parameter-space locations for each individual source population (see Appendix \ref{app:simdetails}). We then calculate  $P(B_i|S_j)$ by binning and normalizing these output samples. Using Bayes' theorem, we write that
\begin{equation}
    {P(S_j|B_i)=\frac{P(B_i|S_j)\,P(S_j)}{P(B_i)}\,.} \label{eq:bayes}
\end{equation}
The probability of finding a sample in a given bin $P(B_i)$ is the sum of the probabilities of being in that bin (given a source population) times the probability of being in that source population. That is, 
\begin{equation}
    {P(B_i)=\sum_j P(B_i|S_j)\,P(S_j)\,.} \label{eq:PBi}
\end{equation}
Note that  $P(B_i|S_j)$ is an output of our simulations. 

\begin{figure*}
    \centering
    \includegraphics[width=\linewidth]{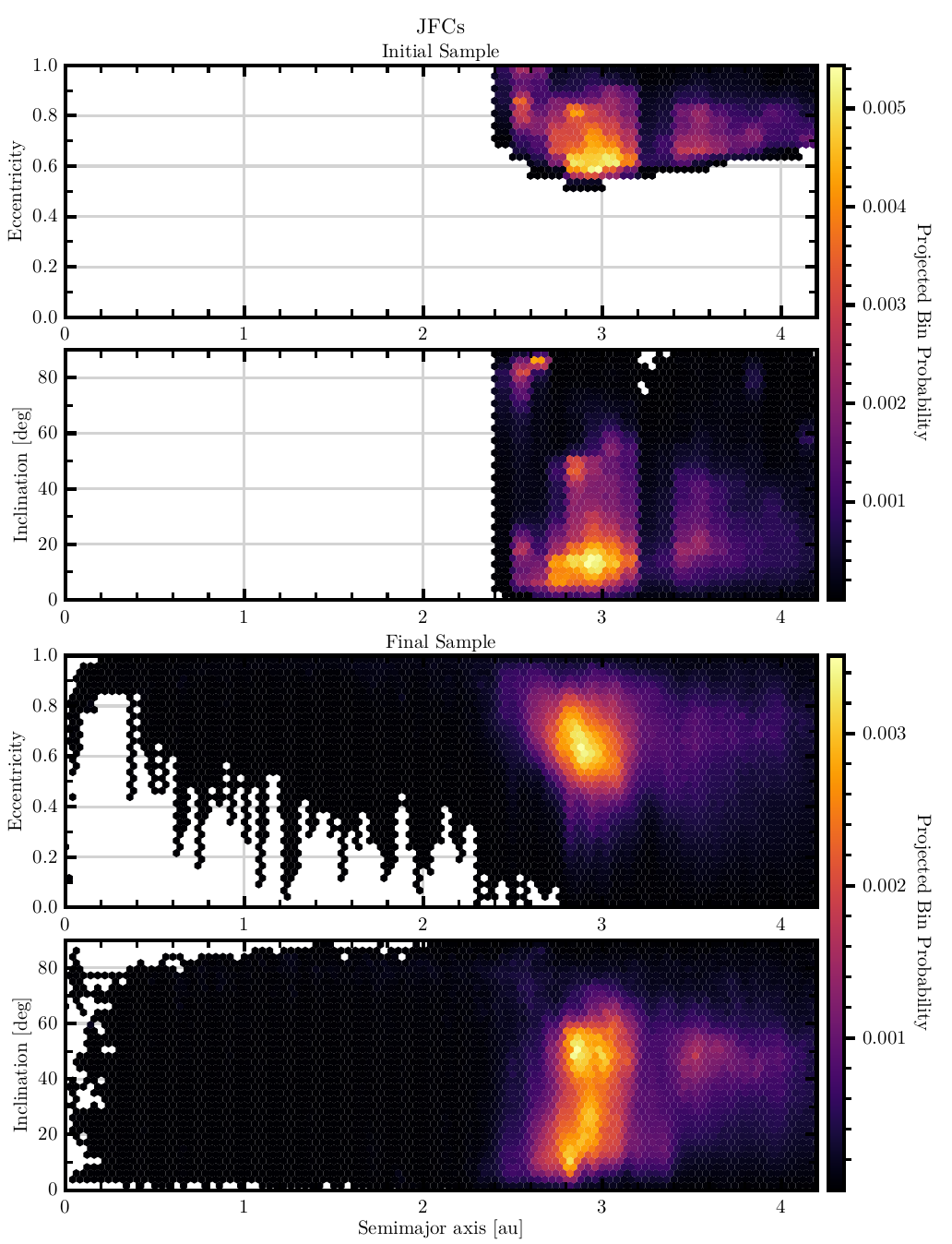}
    \caption{ {The initial and final population distribution for the JFC source, using NEOMOD and \texttt{REBOUNDx}. The colorbar shows the probability of being found in the given projected bin.} }
    \vspace{-10pt}
    \label{fig:sim1}
\end{figure*}

\begin{figure*}
    \centering
    \includegraphics{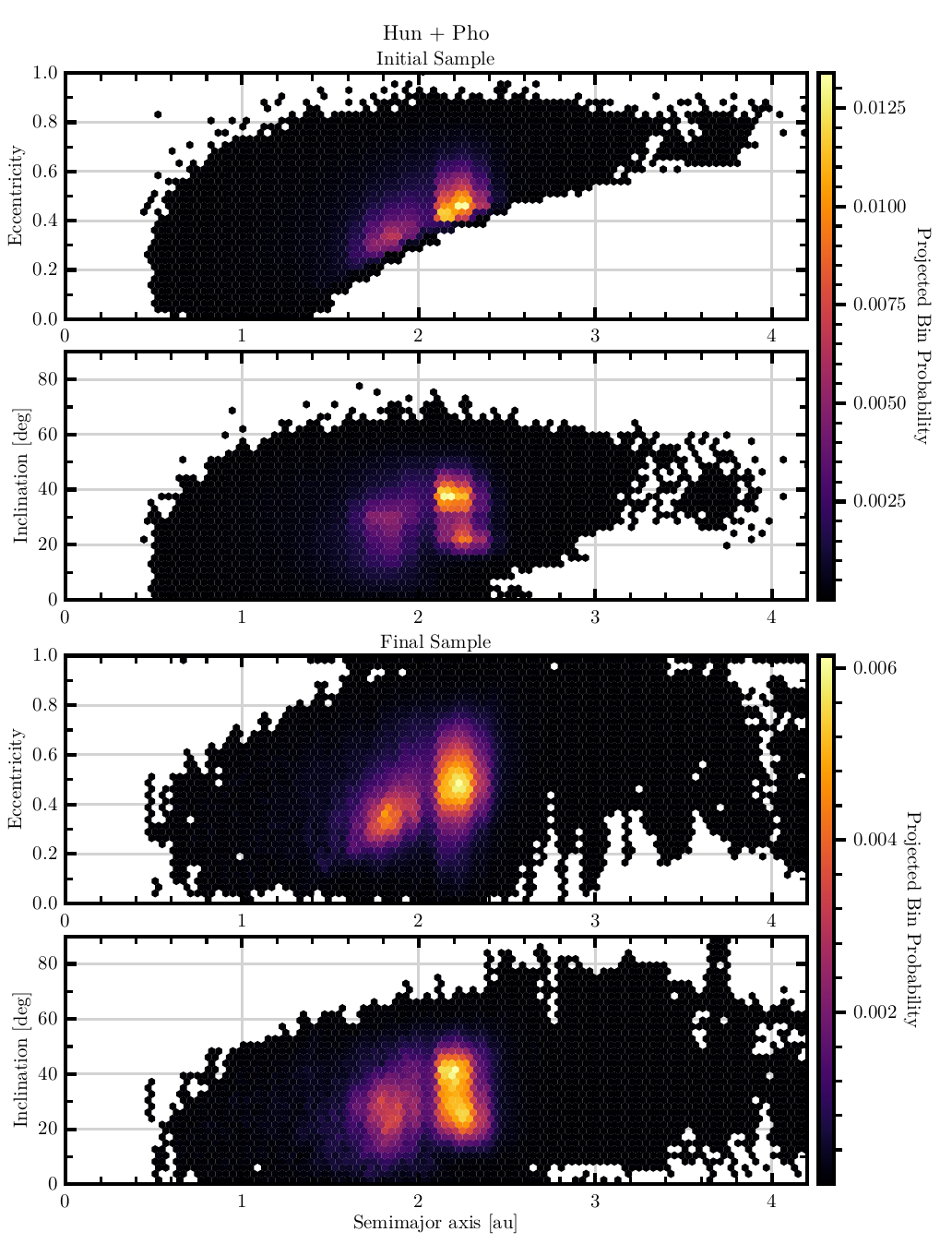}
    \caption{ {The initial and final population distribution for the high-inclination Hungarias and Phocaeas, using NEOMOD and \texttt{REBOUNDx}. The colorbar shows the probability of being found in the given projected bin.} }
    \vspace{-10pt}
    \label{fig:sim2}
\end{figure*}

\begin{figure*}
    \centering
    \includegraphics{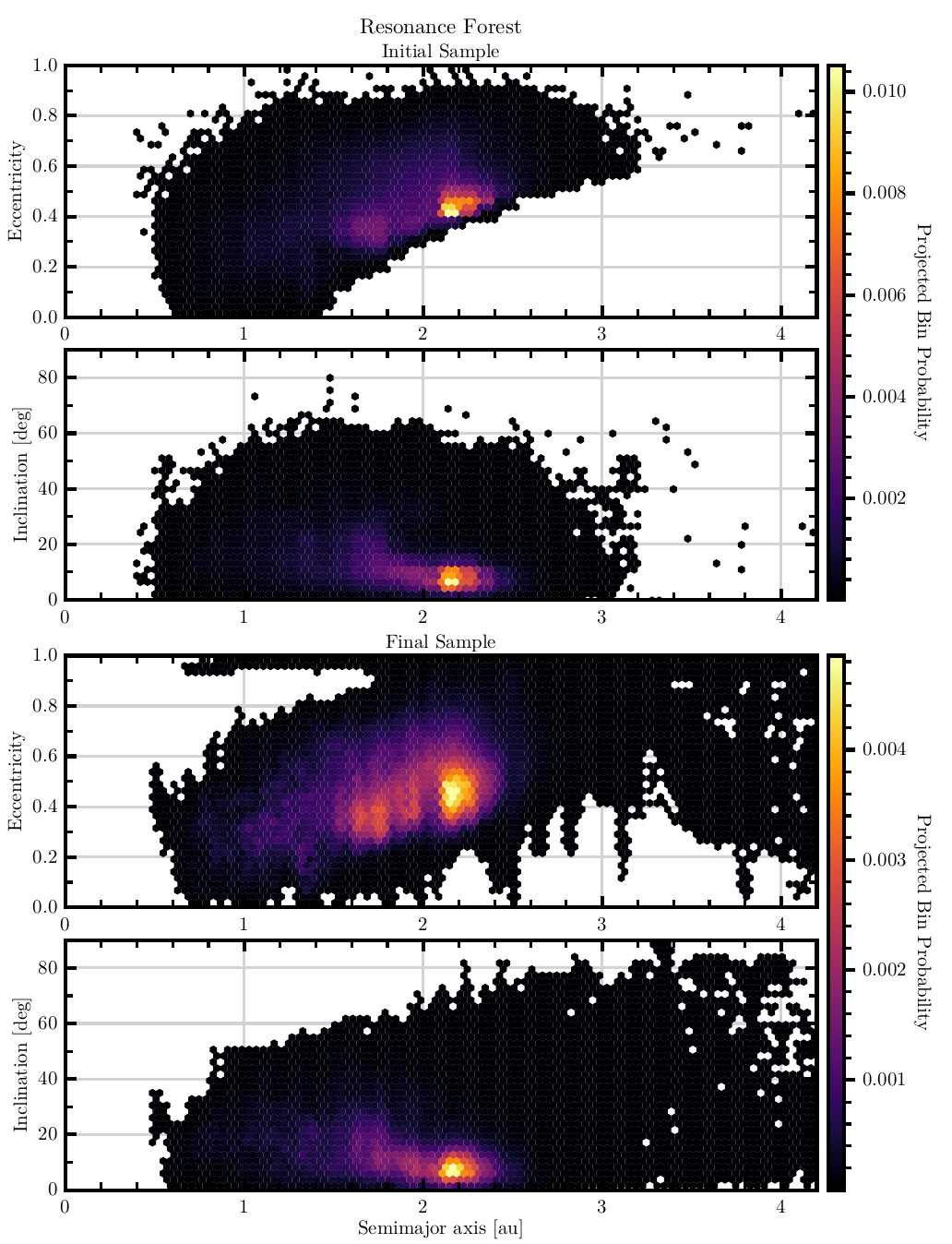}
    \caption{ {The initial and final population distribution for the weak resonance forest source, using NEOMOD and \texttt{REBOUNDx}. The colorbar shows the probability of being found in the given projected bin.} }
    \vspace{-10pt}
    \label{fig:sim3}
\end{figure*}

\begin{figure*}
    \centering
    \includegraphics{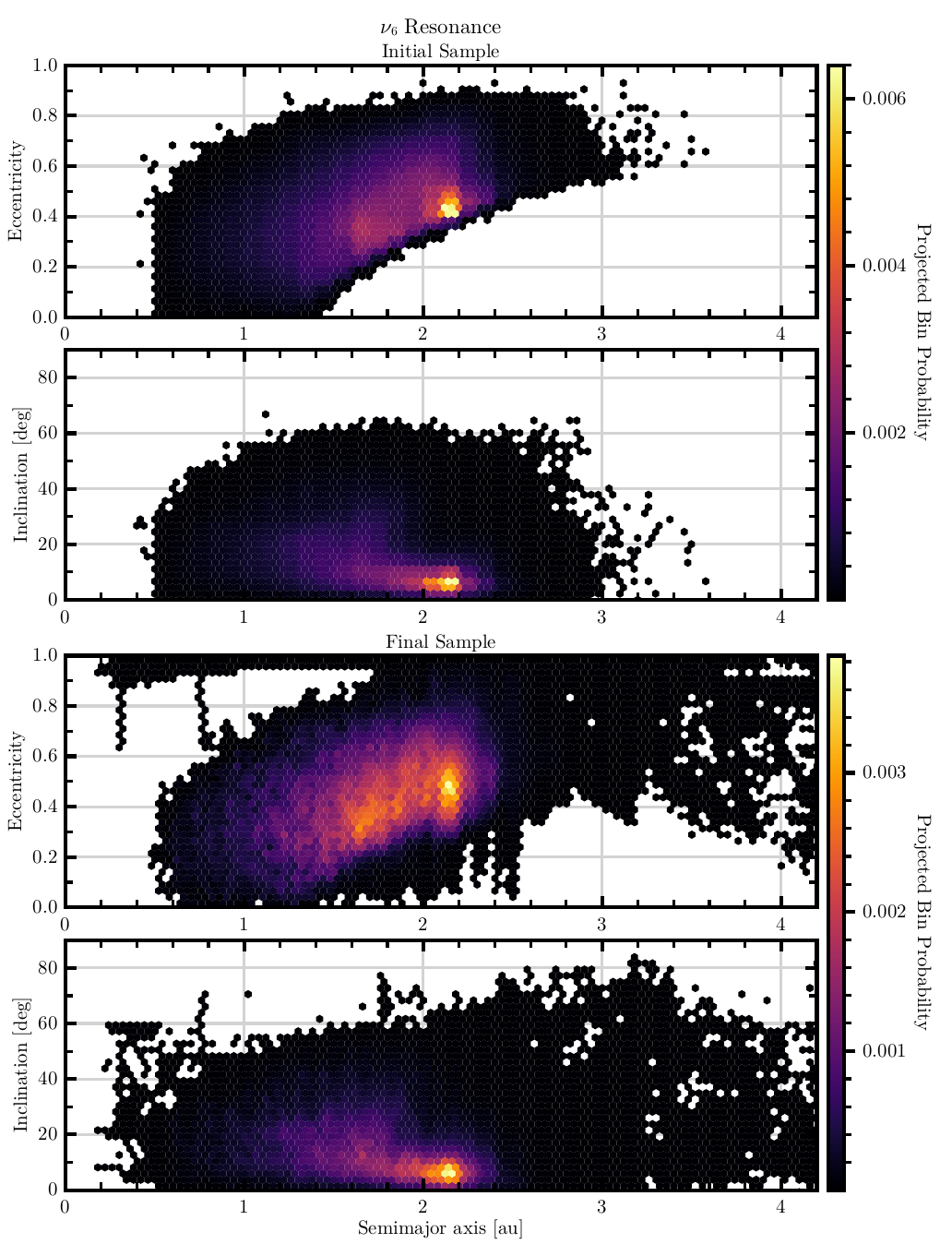}
    \caption{ {The initial and final population distribution for the $\nu_6$ resonance source, using NEOMOD and \texttt{REBOUNDx}. The colorbar shows the probability of being found in the given projected bin.} }
    \vspace{-10pt}
    \label{fig:sim4}
\end{figure*}

\begin{figure*}
    \centering
    \includegraphics{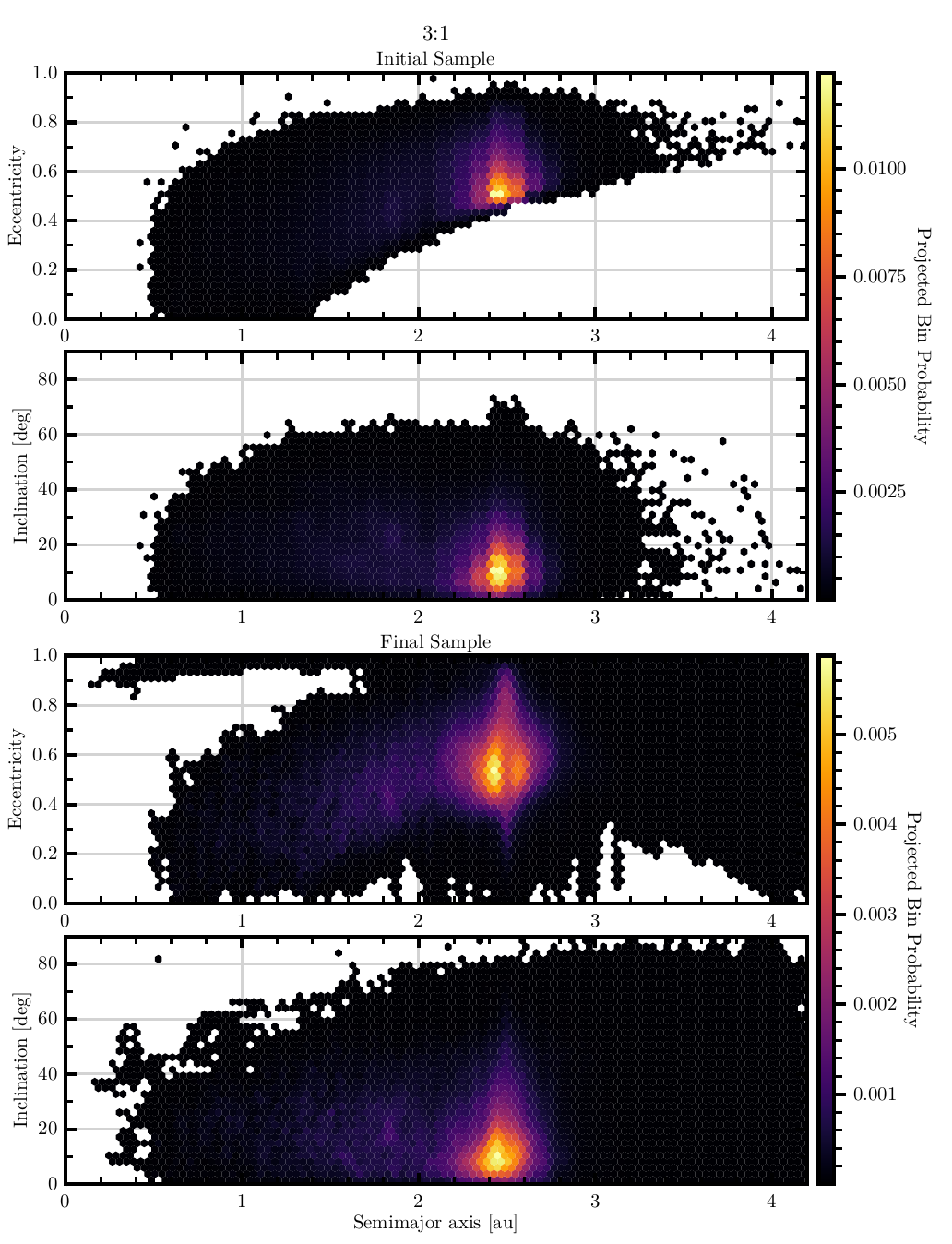}
    \caption{ {The initial and final population distribution for the 3:1 resonance source, using NEOMOD and \texttt{REBOUNDx}. The colorbar shows the probability of being found in the given projected bin.} }
    \vspace{-10pt}
    \label{fig:sim5}
\end{figure*}

\begin{figure*}
    \centering
    \includegraphics{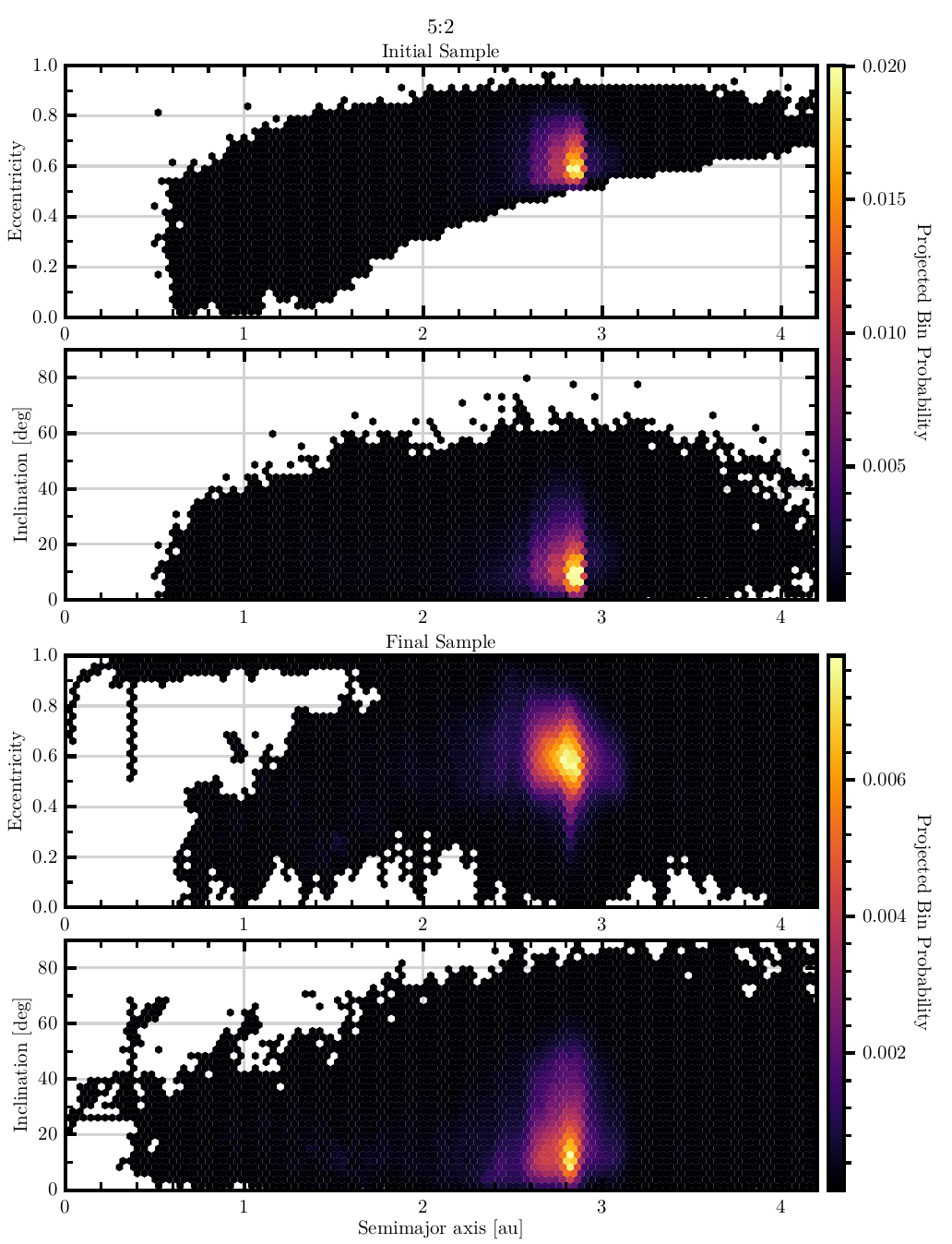}
    \caption{ {The initial and final population distribution for the 5:2 resonance source, using NEOMOD and \texttt{REBOUNDx}. The colorbar shows the probability of being found in the given projected bin.} }
    \vspace{-10pt}
    \label{fig:sim6}
\end{figure*}

\begin{figure*}
    \centering
    \includegraphics{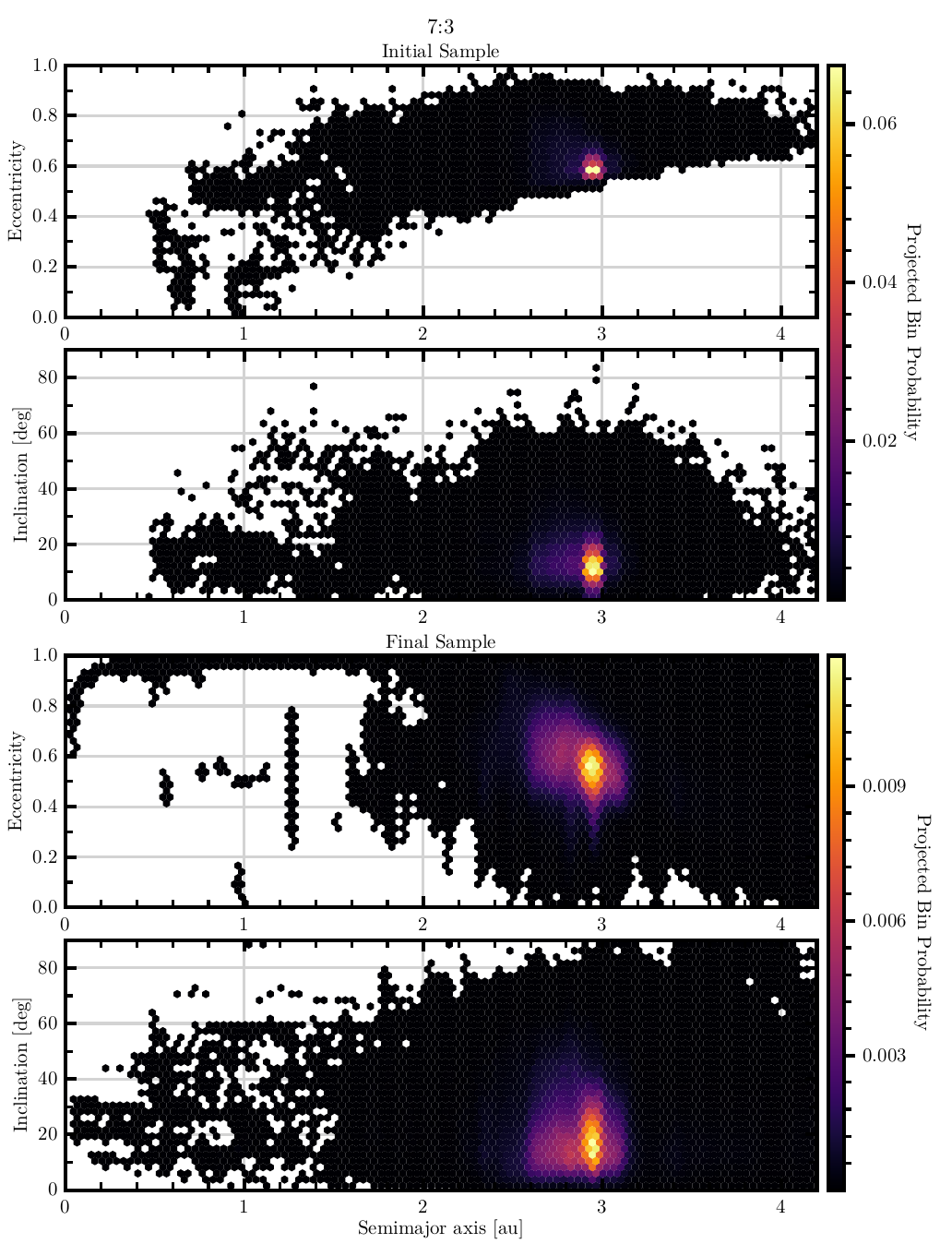}
    \caption{ {The initial and final population distribution for the 7:3 resonance source, using NEOMOD and \texttt{REBOUNDx}. The colorbar shows the probability of being found in the given projected bin.} }
    \vspace{-10pt}
    \label{fig:sim7}
\end{figure*}

\begin{figure*}
    \centering
    \includegraphics{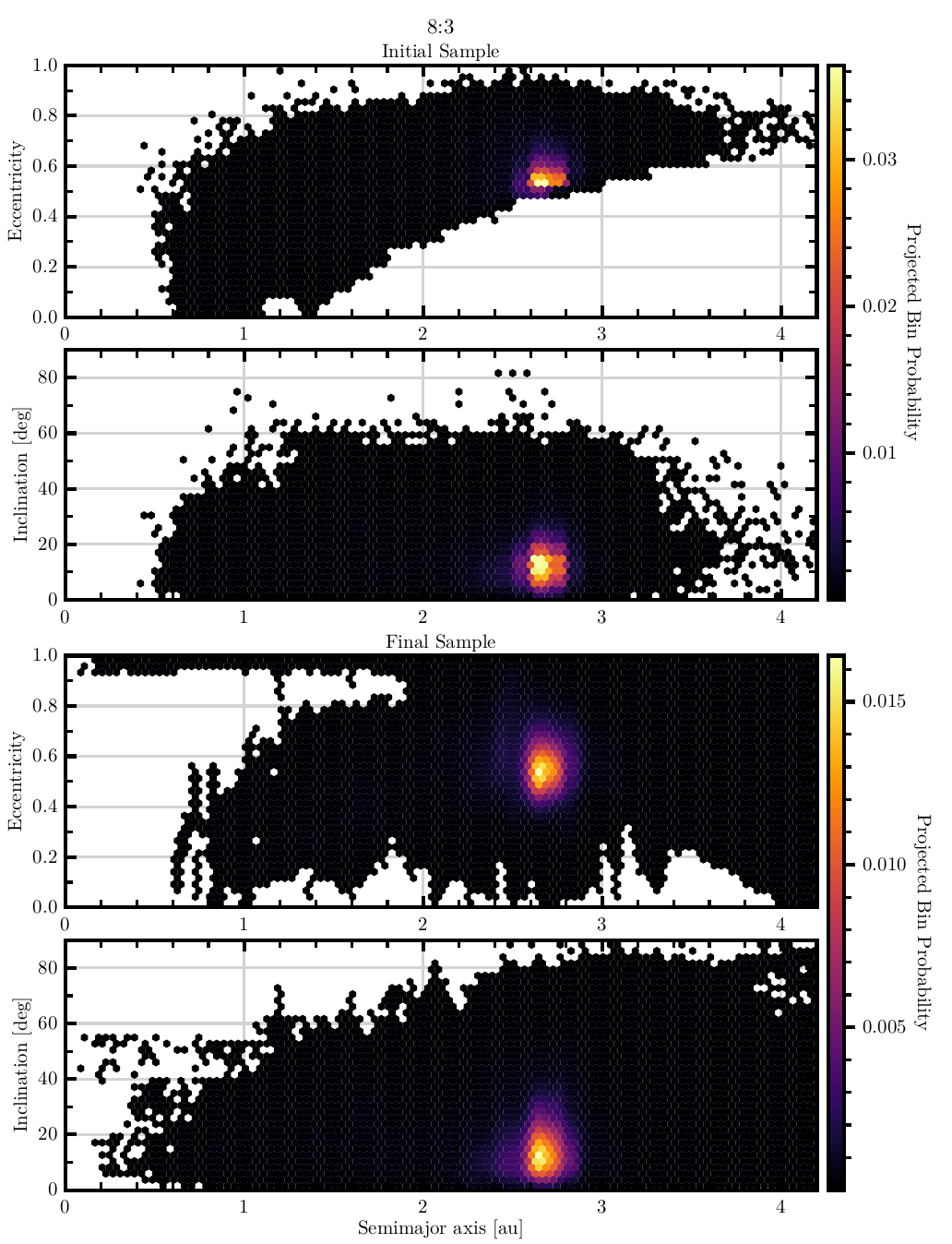}
    \caption{ {The initial and final population distribution for the 8:3 resonance source, using NEOMOD and \texttt{REBOUNDx}. The colorbar shows the probability of being found in the given projected bin.} }
    \vspace{-10pt}
    \label{fig:sim8}
\end{figure*}

\begin{figure*}
    \centering
    \includegraphics{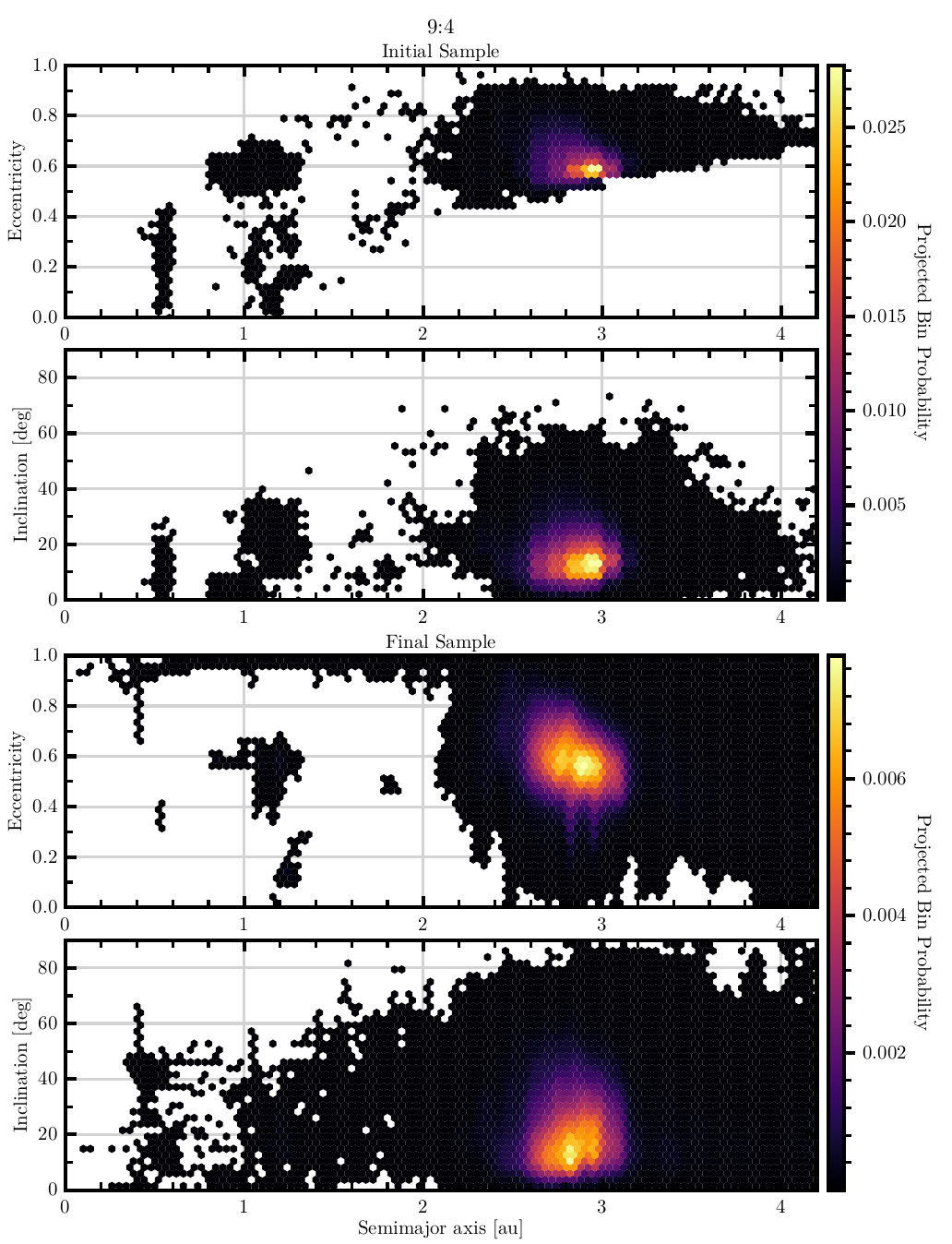}
    \caption{ {The initial and final population distribution for the 9:4 resonance source, using NEOMOD and \texttt{REBOUNDx}. The colorbar shows the probability of being found in the given projected bin.} }
    \vspace{-10pt}
    \label{fig:sim9}
\end{figure*}

\begin{figure*}
    \centering
    \includegraphics{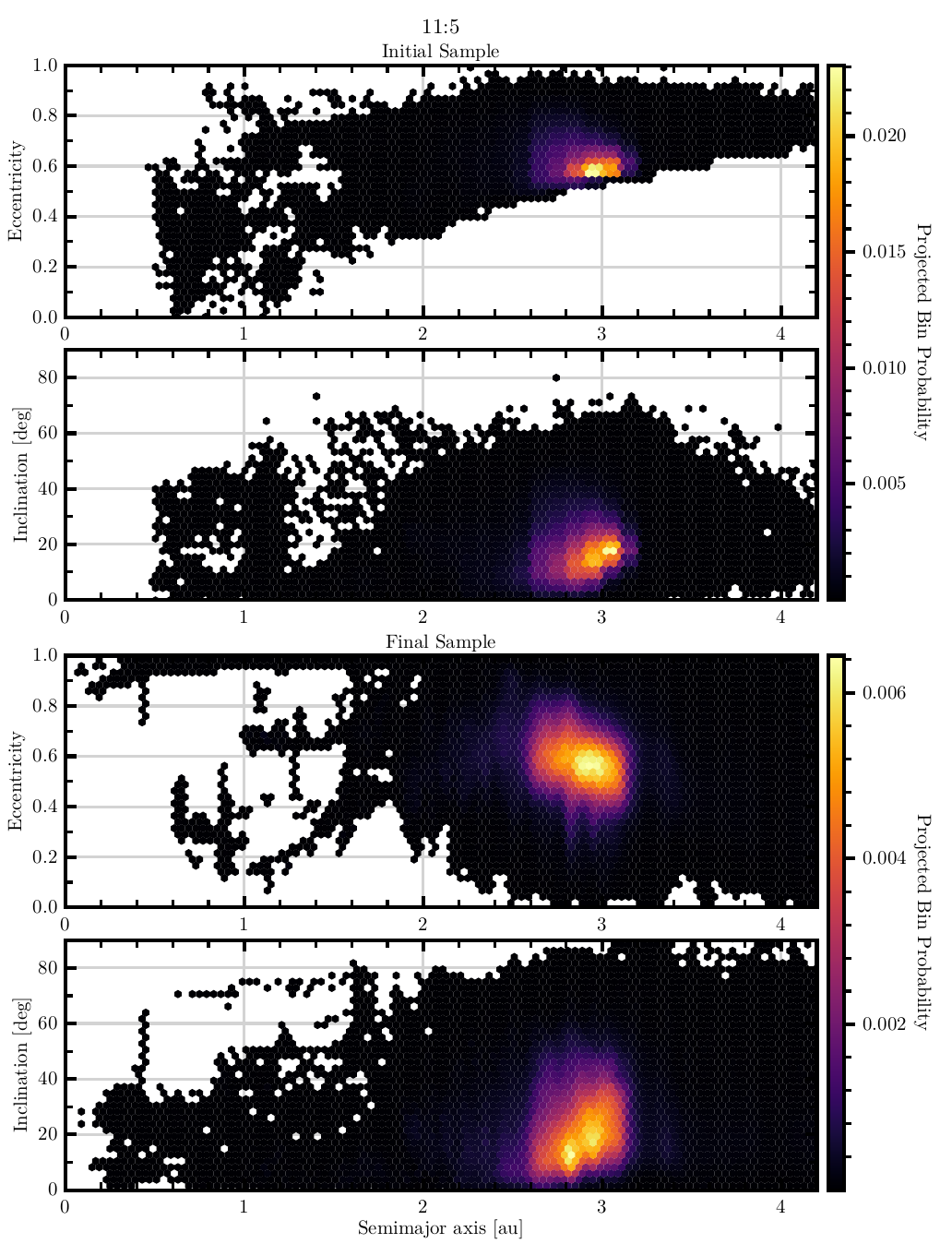}
    \caption{ {The initial and final population distribution for the 11:5 resonance source, using NEOMOD and \texttt{REBOUNDx}. The colorbar shows the probability of being found in the given projected bin.} }
    \vspace{-10pt}
    \label{fig:sim10}
\end{figure*}

\begin{figure*}
    \centering
    \includegraphics{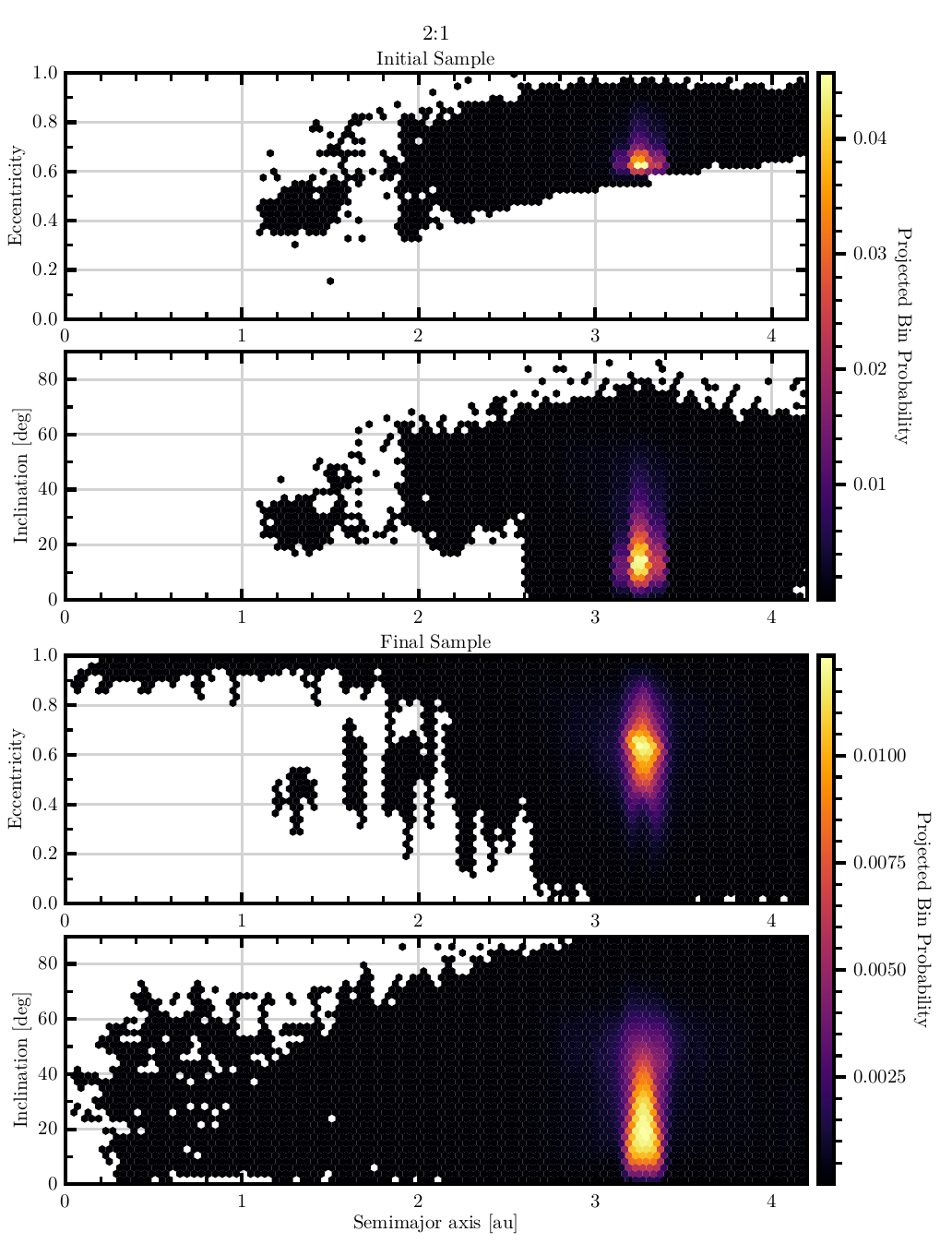}
    \caption{ {The initial and final population distribution for the 2:1 resonance source, using NEOMOD and \texttt{REBOUNDx}. The colorbar shows the probability of being found in the given projected bin.} }
    \vspace{-10pt}
    \label{fig:sim11}
\end{figure*}

\begin{figure*}
    \centering
    \includegraphics{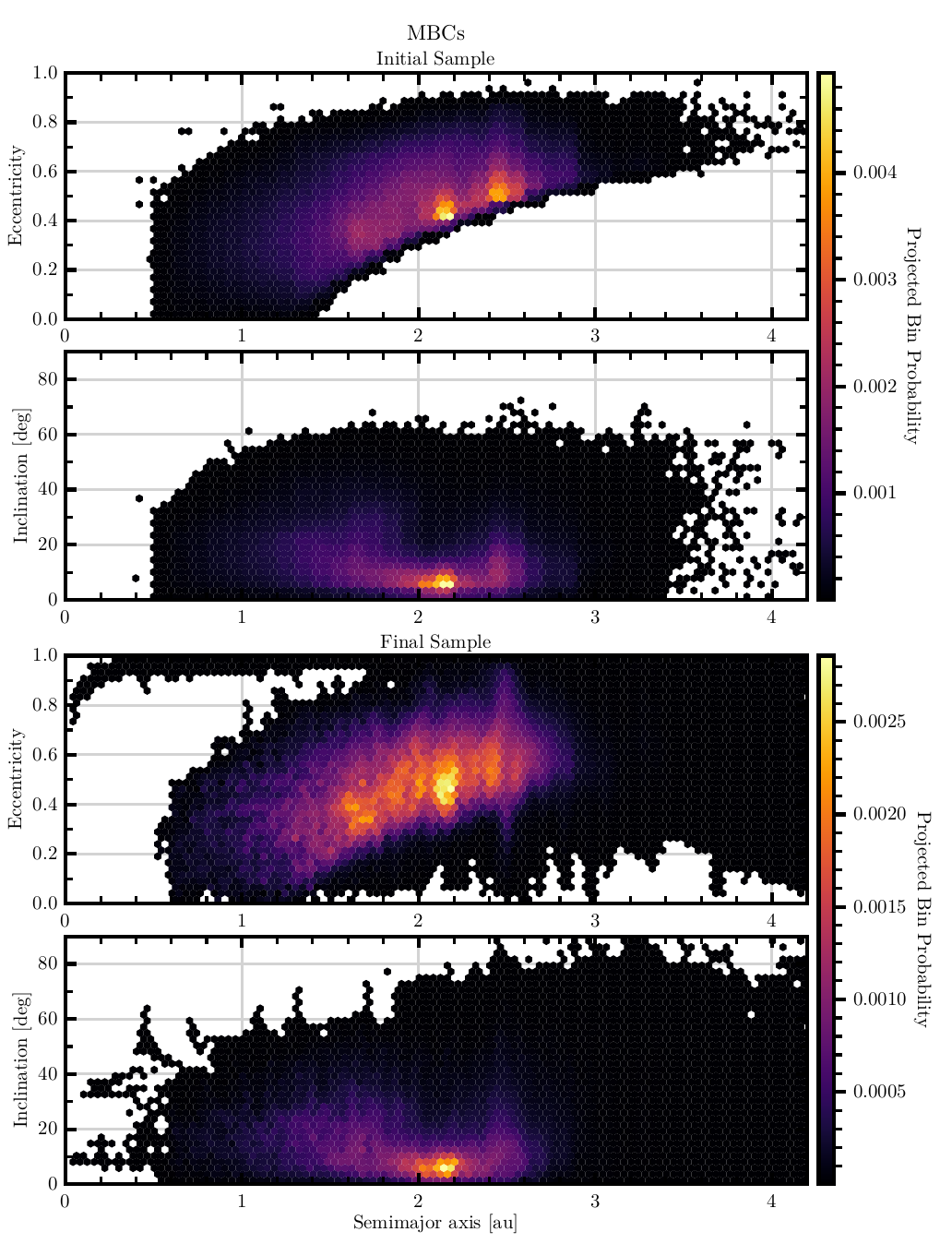}
    \caption{ {The initial and final population distribution for a sample drawn from all main belt objects, using NEOMOD and \texttt{REBOUNDx}. The colorbar shows the probability of being found in the given projected bin.} }
    \vspace{-10pt}
    \label{fig:sim12}
\end{figure*}

The NEOMOD data file provides $P_{\rm NMD}(S_j|B_i)$ and $P_{\rm NMD}(B_i)$. Analogously to Eq. \eqref{eq:PBi}, we therefore find that 
\begin{equation}
    P(S_j)=\sum_i P_{\rm NMD}(S_j|B_i)\,P_{\rm NMD}(B_i)\,.\label{eq:PSj}
\end{equation}
These values of $P(S_j)$ are given in Table \ref{table:PSjs}. Note that the $P(S_j)$'s calculated in Eq. \eqref{eq:PSj} are based on the NEOMOD model, which has different population distributions from our data (see Figs. \ref{fig:sim1}--\ref{fig:sim12}).

We use Eq. \eqref{eq:bayes} to calculate the probability that a given dark comet is from a given source. These values are reported in Table \ref{table:dcprobs}. The limits and bins on our calculations are identical to that of NEOMOD --- 42 bins in semimajor axis from 0 to 4.2 au, 25 bins in eccentricity from 0 to 1, and 22 bins in inclination from 0 to 90$^\circ$. 

Note that the values shown in Table \ref{table:dcprobs} are only order-of-magnitude estimates as a result of several limitations to this methodology. First, we use the normalization from NEOMOD, rather than computing our own normalization by comparing to observational data. Second, in our calculations we marginalize over the absolute magnitude, reducing a dimension in our sample. Third, our probability calculations do not account for any morphological properties of the dark comets, which may further constraint the dark comets' source population. 

\end{document}